%% file: main.tex
\documentclass{article}

\usepackage[preprint]{neurips_2026}

\usepackage[utf8]{inputenc}
\usepackage[T1]{fontenc}
\usepackage{hyperref}
\usepackage{url}
\usepackage{booktabs}
\usepackage{amsfonts}
\usepackage{amsmath}
\usepackage{amssymb}
\usepackage{nicefrac}
\usepackage{microtype}
\usepackage{xcolor}
\usepackage{graphicx}

\usepackage{footmisc}  
\usepackage{tabularx}
\usepackage{multirow}
\usepackage{colortbl}
\usepackage{pifont}  
\usepackage{tikz}
\usepackage{latexsym}

\newcommand{\todo}[1]{}
\newcommand{\note}[1]{}

\title{A Survey of Full-Duplex Spoken Dialogue Systems:\\
       Architectural Hierarchy, Interaction Ontology, and Decision State Machine}

\author{%
  \mdseries
  Jingyu Lu$^{1,2}$ \quad Yuhan Wang$^{1,2}$ \quad Jianming Luo$^{1}$ \quad Yifu Chen$^{1}$ \quad Tianle Liang$^{1}$ \\
  Shengpeng Ji$^{3}$ \quad Ziyue Jiang$^{1}$ \quad Xiaoda Yang$^{1}$ \quad Yu Zhang$^{4}$ \quad Xize Cheng$^{1}$ \\
  Chenyuhao Wen$^{1}$ \quad Changhao Pan$^{1}$ \quad Haoxiao Wang$^{1}$ \quad Chen Ye$^{2}$ \quad Jian Wu$^{2}$ \\
  Xiaoxi Jiang$^{2}$ \quad Guanjun Jiang$^{2}$ \quad Zhou Zhao$^{1}$ \\[0.6em]
  \normalsize
  $^{1}$Zhejiang University \quad
  $^{2}$Qwen Business Unit, Alibaba \quad
  $^{3}$HunYuan Team, Tencent \quad
  $^{4}$ByteDance \\[0.3em]
  \url{https://github.com/DuplexLM/DuplexSurvey}
}

\begin{document}

\maketitle

\begin{abstract}
More than a dozen spoken dialogue systems have recently claimed to be ``full-duplex,'' yet the term has been used to describe substantially different capabilities. Existing surveys collapse them onto a single axis (cascaded/end-to-end, or engineered/learned) and miss the distinctions that matter most for builders. We argue that much of this ambiguity is taxonomical: current terminology does not specify where duplex decisions are made, which interaction types are supported, or how a system behaves moment by moment. This paper introduces three complementary frameworks: \textbf{(i)} an \textbf{L0--L3 Architectural Hierarchy} that locates where duplex decisions are made; \textbf{(ii)} a \textbf{$T \times I \times R$ Interaction Ontology} that specifies the temporal relation, user intent, and required system response for each interaction; and \textbf{(iii)} a \textbf{Decision State Machine} (\textsc{Idle}/\textsc{Listen}/\textsc{Speak}/\textsc{Wait}/\textsc{Dual}) that describes how systems move between states. Across published systems and benchmarks, our audit documents a realization gap: although many architectures can in principle operate in full-duplex states, their observed behavior remains constrained by the interaction patterns represented in training and evaluation. We point to the limited public training-data coverage relative to the (largely undisclosed) industrial corpora, together with the still-unrealized goal of L3 representation-level modeling, as the key frontiers for future research on full-duplex dialogue. The related material is available at \href{https://github.com/DuplexLM/DuplexSurvey}{this https url}.
\end{abstract}


\input{sections/01_introduction}            
\input{appendix/A_foundations}              
\input{sections/02_history}                 
\input{appendix/B_pre_fd_era}               
\input{sections/03_hierarchy_audit}         
\input{appendix/C_systems}                  
\input{sections/04_ontology}                
\input{appendix/C_cells}                    
\input{sections/05_state_machine}           
\input{appendix/F_state_machine_proofs}     
\input{sections/06_data}                    
\input{appendix/D_data}                     
\input{sections/07_evaluation}              
\input{appendix/E_evaluation}               
\input{sections/08_frontiers_conclusion}    

\bibliographystyle{unsrtnat}
\bibliography{custom}

\end{document}

%% file: sections/01_introduction.tex
\section{Introduction}
\label{sec:intro}

Full-duplex spoken dialogue --- concurrent speaking and listening with overlap, backchannels, and barge-ins --- is not a strictly new capability. Pre-LLM industrial pipelines such as Google Duplex~\citep{leviathan2018google}, Ant Group~\citep{jin2021duplex}, and Alibaba DAMO~\citep{lin2022duplex} had already deployed it at production scale by 2022. What changed with OpenAI's GPT-4o demonstration of May 2024\footnote{\url{https://openai.com/index/hello-gpt-4o/}} was the architectural register: an end-to-end speech-to-speech model interruptible mid-utterance turned full-duplex into a baseline product expectation, and within two years more than a dozen new systems --- Moshi~\citep{defossez2024moshi}, MinMo~\citep{chen2025minmo}, OmniFlatten~\citep{zhang2025omniflatten}, SyncLLM~\citep{veluri2024beyond}, LSLM~\citep{ma2025language}, FireRedChat~\citep{chen2025fireredchat}, and many others --- have shipped under this banner.

These systems share the banner but differ substantially in what they do. One implements full-duplex as a modular controller that yields the floor upon a detected interruption; another encodes simultaneous user and assistant audio in parallel token streams of the same decoder; a third runs a small predictor on the LLM's hidden state and produces speech through a downstream module. Their interaction coverage, latency profiles, and failure modes overlap only partially, yet existing surveys~\citep{ji2024wavchat,chen2025turn} tend to collapse the design space onto a single axis (cascaded vs end-to-end, or engineered vs learned synchronization), and consequently cannot articulate \emph{where} a duplex decision is made in the model stack, \emph{which} interaction types a system actually handles, or \emph{how} its behavior evolves moment by moment.

A second observation, visible only once these three dimensions are read together, is that full-duplex behavior tends to appear as a joint product of architectural reach, training-data coverage, and per-moment policy rather than a feature contributed by any one of them: nearly identical L2 decoders can serve very different interaction cells when their training corpora differ, and systems trained on the same data can fail differently when the duplex decision is wired into different layers of the stack. We therefore organize the survey around this composite view, providing three analytical frameworks that make each factor explicit and three frameworks-grounded audits that put them to work on the published evidence.

The first framework, an \textbf{L0--L3 Architectural Hierarchy} (\S\ref{sec:hierarchy-audit}), locates the duplex decision in the model stack along four layers --- external module (L0), hidden-state predictor (L1), token-level synchronization (L2), or shared latent representation (L3) --- with L1 emerging as a recurring structural pattern across product goals (full-duplex MinMo / Freeze-Omni and non-FD streaming Qwen Thinker--Talker / Step-Audio R1.1 share the same shape) while L2 has fragmented into a broader industrial design space; L3 remains concretely open. The second, a \textbf{T\,$\times$\,I\,$\times$\,R Interaction Ontology} (\S\ref{sec:ontology}), characterizes a full-duplex moment as a triple over a temporal relation, a user intent, and a required system response, and singles out six acid-test cells that, jointly, distinguish a genuine full-duplex system from a half-duplex one. The third, a \textbf{Decision State Machine} (\S\ref{sec:state-machine}) with five states ($\{\textsc{Idle}, \textsc{Listen}, \textsc{Speak}, \textsc{Wait}, \textsc{Dual}\}$) and eleven transitions, extends the FSM lineage of \citet{wang2024full} and \citet{liao2025flexduo} and couples every transition to a specific ontology cell, turning ``what is the system doing at this instant?'' into a structured question.

We then use the three frameworks to audit the published evidence along three matching dimensions. The \textbf{architecture audit} (\S\ref{sec:hierarchy-audit}, Table~\ref{tab:audit}) places the leading frontier systems on the L0--L3 axis with their reachable FSM states and ontology-cell coverage. The \textbf{training-data audit} (\S\ref{sec:data}) traces cross-system variance at L0--L2 to the cell coverage of public corpora and supports a data-bottleneck thesis with three independent pieces of evidence, including Moshi's documented use of Fisher for full-duplex fine-tuning and MinMo's 4000\,h dialogue mix. The \textbf{evaluation audit} (\S\ref{sec:evaluation}) introduces the \emph{realization gap} between architectural capacity and demonstrated behavior as a diagnostic frame, uses it to clarify what current benchmarks actually test, and disambiguates the two simultaneously active FDB and FD-Bench series whose similar names have caused recurring confusion in the literature. Finally, we close (\S\ref{sec:frontiers-conclusion}) with three concrete L3 architectural hypotheses --- continuous-latent autoregressive streaming, JEPA-style dual-stream prediction, and world-model-conditioned dialogue --- each anchored in recent text-vision continuous-latent and JEPA results. Figure~\ref{fig:paper-overview} gives a single-page navigation of the nine chapters and the systems, corpora, and benchmarks discussed in each.

\paragraph{Survey scope.} We structure the empirical audits around three categories of systems: (i)~primary targets, comprising post-GPT-4o systems that explicitly claim full-duplex behavior; (ii)~historical baselines, represented by pre-2024 industrial L0 pipelines; and (iii)~taxonomy references, such as adjacent streaming speech-LMs (e.g., GLM-4-Voice, Qwen-Omni series), which are included to contextualize the design space but are excluded from per-cell audits. Models restricted to TTS-only, ASR-only, voice-cloning, or non-dialogue audio generation are out of scope.

\input{figures/figure_paper_overview}

%% file: figures/figure_paper_overview.tex
\begin{figure}[!htbp]
\centering
\includegraphics[width=\textwidth, keepaspectratio]{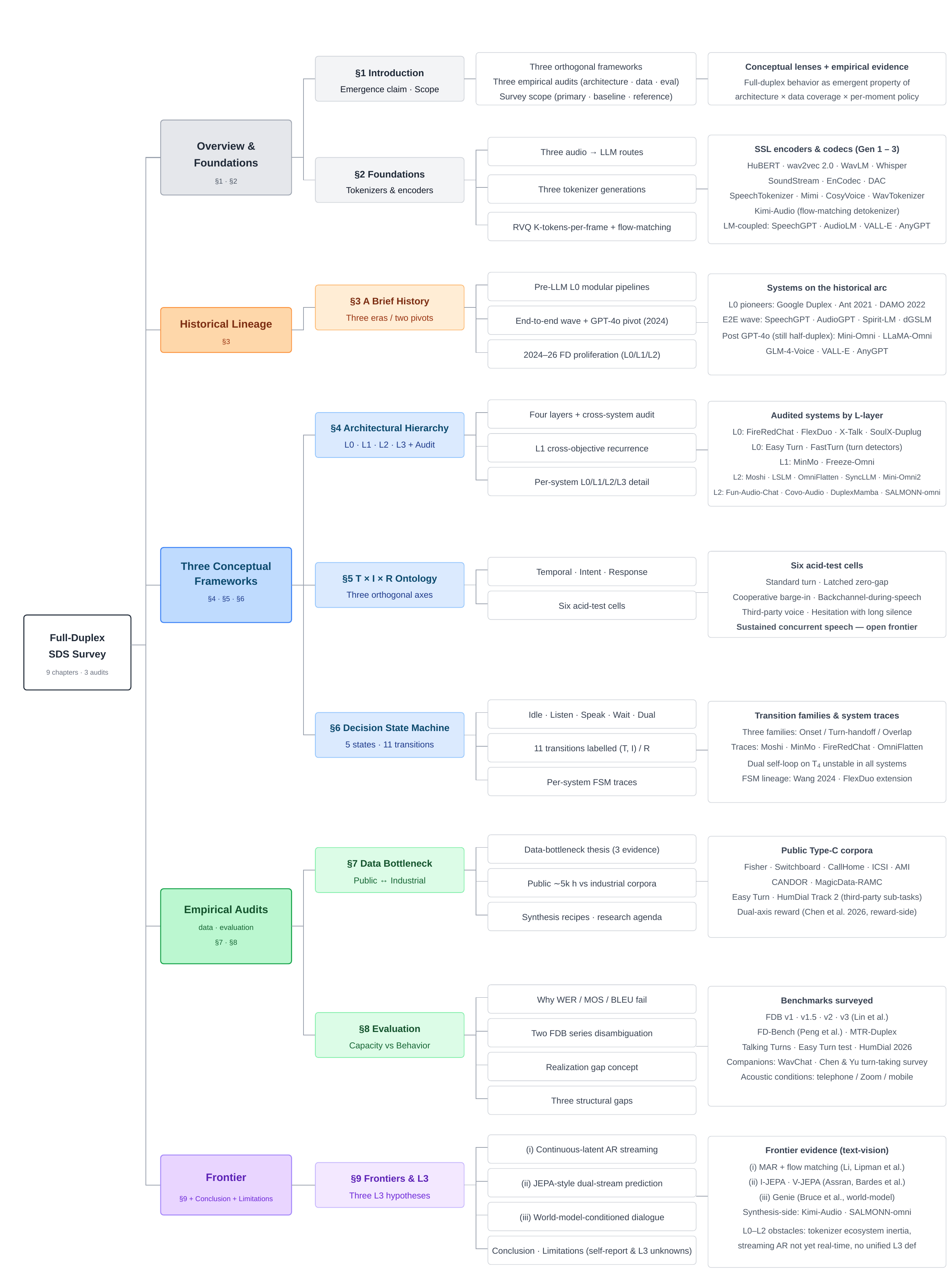}
\caption{Overview of the survey structure. Nine chapters are grouped into four reading phases: \textbf{background} (\S\ref{sec:intro}--\S\ref{sec:history}), \textbf{three conceptual frameworks} (\S\ref{sec:hierarchy-audit}--\S\ref{sec:state-machine}: L0--L3 Architectural Hierarchy, T\,$\times$\,I\,$\times$\,R Interaction Ontology, and the five-state Decision State Machine), \textbf{empirical audits} (\S\ref{sec:data}--\S\ref{sec:evaluation}: a training-data audit and an evaluation audit organized under a unified realization-gap reading), and the \textbf{forward-looking frontier} (\S\ref{sec:frontiers-conclusion}, three L3 architectural hypotheses). Each row gives the chapter title, its key content, and the exemplar systems / corpora / benchmarks discussed.}
\label{fig:paper-overview}
\vspace{-8pt}
\end{figure}

%% file: appendix/A_foundations.tex
\section{Foundations: Audio Tokenizer and Encoder Lineage}
\label{sec:foundations}
\label{app:foundations}
\label{app:foundations}

The L2 token-level full-duplex systems audited in \S\ref{app:systems} all share an upstream dependency: they assume audio has already been transformed into a representation a Transformer language model can consume. The form of that representation --- which tokenizer, with what frame rate, how many codebooks, and whether the first codebook is shared with text --- determines what the L2 architecture can express. This section provides the minimum tokenizer and encoder background needed to read the per-system breakdowns in \S\ref{app:systems} and the sub-design discussions in \S\ref{sec:ha:audit}. It is not a tokenizer survey; it identifies the three pieces of upstream infrastructure on which the rest of the paper rests.

\subsection{Three Routes from Audio to LLM}
\label{app:found:routes}

Three structurally distinct routes have been used to deliver speech into a Transformer LLM, and each maps onto a recognizable line of L2 systems:

\begin{itemize}
  \setlength{\itemsep}{1pt}
  \item \textbf{Discrete audio tokens.} The waveform is quantized into a sequence of integer tokens that the LLM can predict the same way it predicts text tokens. This route is the most-traveled because it reuses the LLM training and inference stack without modification; it is what underlies Mini-Omni~\citep{xie2024mini}, Moshi~\citep{defossez2024moshi}, OmniFlatten~\citep{zhang2025omniflatten}, GLM-4-Voice~\citep{zeng2024glm}, and SyncLLM~\citep{veluri2024beyond}.
  \item \textbf{Continuous audio embeddings.} A streaming speech encoder (e.g., a Whisper-encoder or a custom SSL encoder) outputs continuous vectors that are aligned to the LLM's hidden-state space via a projector. LLaMA-Omni~\citep{fang2025llama} is the canonical example; the route preserves more acoustic information than tokenization but requires special handling on the input side of the LLM.
  \item \textbf{Hybrid text-leads / audio-trails.} A text token stream drives generation, with audio tokens emitted in parallel or shortly thereafter conditioned on the text. Moshi's Inner Monologue~\citep{defossez2024moshi} is the load-bearing instance: the text track provides the semantic backbone and the audio tracks attach acoustic realization to it.
\end{itemize}

Routes (i) and (iii) together account for essentially all L2 systems in \S\ref{app:systems}; route (ii) is the natural home for the codec-free designs (SALMONN-omni~\citep{yu2026salmonn} brought this route into a full-duplex setting in 2026).

\subsection{Three Generations of Audio Tokenizers}
\label{app:found:tokenizers}

Discrete tokenization for speech has gone through three generations, each defined by what it optimizes for and what it loses. The L2 explosion of 2024 is downstream of the third generation reaching engineering maturity.

\paragraph{Generation~1 --- semantic units.} HuBERT~\citep{hsu2021hubert}, wav2vec~2.0~\citep{baevski2020wav2vec}, w2v-BERT~\citep{chung2021w2v}, and WavLM~\citep{chen2022wavlm} (2020--2021) trained self-supervised speech encoders, with discrete tokens obtained via $k$-means clustering over hidden states at roughly 25--50\,Hz. Whisper~\citep{radford2023robust} is the dominant supervised counterpart used as a streaming encoder for the continuous-embedding route (\S\ref{app:found:routes}, route (ii)). Property: semantically faithful (the unit sequence preserves what was said) but acoustically lossy (reconstruction loses speaker identity and prosody). Suitable for understanding tasks; unsuitable as the sole representation when the system must also speak naturally. SpeechGPT~\citep{zhang2023speechgpt} and AnyGPT~\citep{zhan2024anygpt} are canonical generation-1 LLM-coupled systems.

\paragraph{Generation~2 --- neural audio codecs.} SoundStream~\citep{zeghidour2021soundstream}, EnCodec~\citep{defossez2022high}, and DAC~\citep{kumar2023high} reframed audio tokenization as a signal-compression problem: an encoder--quantizer--decoder is trained with residual vector quantization (RVQ) to compress the waveform to roughly 1--3\,kbps with high reconstruction fidelity. A typical configuration is 75\,Hz frame rate with 8 RVQ codebooks per frame. AudioLM~\citep{borsos2023audiolm} and VALL-E~\citep{wang2023neural} are the canonical generation-2 LM-coupled systems on the generation side. Property: acoustically faithful but semantically weak --- the eight codebooks of a single frame have no canonical ordering by importance, so an LLM trained on the flattened token stream learns acoustic detail before semantic content, hurting downstream language behavior.

\paragraph{Generation~3 --- semantic--acoustic fused.} SpeechTokenizer~\citep{zhang2308speechtokenizer}, Mimi (the codec used by Moshi~\citep{defossez2024moshi}), and the CosyVoice tokenizer~\citep{du2024cosyvoice} added an explicit semantic-then-acoustic ordering to the RVQ stack: the first codebook is distilled or constrained to carry semantic content (in Mimi's case, distilled from WavLM~\citep{chen2022wavlm}), and the remaining codebooks carry acoustic residuals. WavTokenizer~\citep{ji2025wavtokenizer} represents the parallel single-codebook line that compresses to similarly low bitrates without the multi-codebook structure. Property: the first codebook of generation~3 behaves approximately like a text token, which lets an LLM read or write speech with the same training pipeline as text. Mimi specifically operates at 12.5\,Hz with 8 codebooks, totalling 1.1\,kbps --- low enough that an autoregressive Transformer can produce real-time speech. Generation~3 is the upstream condition for Moshi and for the broader L2 token-level wave of 2024; without it, the L2 row of the audit table in \S\ref{sec:ha:audit} would not exist.

\subsection{The K-Tokens-per-Frame Problem}
\label{app:found:rvq}

A direct consequence of generations~2 and~3 both using RVQ is that every audio frame produces $K$ tokens rather than one. For Mimi, $K = 8$. A naive autoregressive treatment serializes the eight codebooks along the time axis, multiplying the sequence length and the inference cost by eight; for ten seconds of audio at 12.5\,Hz, this turns 125 frames into 1000 tokens. The L2 sub-designs surveyed in \S\ref{sec:ha:audit} are, to a first approximation, different answers to ``how should the $K$ codebooks of each frame be arranged in the autoregressive sequence?'' Moshi's multi-stream RQ-Transformer predicts all $K$ codebooks at the same time step using a small inner Transformer; OmniFlatten flattens the codebooks along time; SyncLLM packs each chunk's tokens into a fixed-length block. These are not architectural curiosities at the LLM level; they are direct responses to the K-tokens-per-frame constraint imposed by the tokenizer.

\subsection{Two Recent Departures}
\label{app:found:departures}

Two 2025--2026 directions broaden the tokenizer assumption space and matter for the L2 design discussion in \S\ref{sec:ha:audit}.

\paragraph{Flow-matching streaming detokenizer.} Kimi-Audio~\citep{ding2025kimi} decouples the LLM input side from the LLM output side: the LLM continues to consume discrete tokens, but the audio synthesis side uses a chunk-wise streaming flow-matching detokenizer rather than an RVQ-decoder. This widens the design space because the LLM no longer has to predict the same token representation it consumes, and naturalness on the output side can be improved without re-training the LLM. The pattern is too new to have settled into a sub-design but already appears in several 2026 system reports.

\paragraph{Codec-free representations.} SALMONN-omni~\citep{yu2026salmonn} drops the audio codec entirely and operates on continuous embeddings throughout, bringing route (ii) of \S\ref{app:found:routes} into the L2 design space. The trade-off is the inverse of generation~3 tokenization: the model retains more acoustic detail at the cost of losing the ``audio as language'' framing that made the LLM training pipeline reusable in the first place. SALMONN-omni's appearance is what makes the L2 row of \S\ref{sec:ha:audit} structurally heterogeneous rather than uniformly token-based.

\subsection{Reading Guide}

For the rest of the paper, three takeaways from this section are load-bearing. \emph{First}, when an L2 system is described as ``token-level full-duplex'', the precise meaning depends on which generation of tokenizer it uses: generation~3 lets the same Transformer carry semantic and acoustic content, which is why the L2 wave clusters in late 2024 rather than earlier. \emph{Second}, the L2 sub-design names used informally in \S\ref{sec:ha:audit} (multi-stream, flatten, chunk-alternation) are upstream-determined by the K-tokens-per-frame constraint, not invented top-down at the LLM level. \emph{Third}, route (ii) (continuous embeddings) and the codec-free designs are not exceptions to the L2 framing of \S\ref{sec:ha:audit} but legitimate L2 systems whose decision point is on the output token stream of a non-RVQ representation. Throughout the per-system breakdowns of \S\ref{app:systems} we use ``L2 token-level'' to mean the decision is at the discrete-output layer, whether that layer is RVQ-coded or continuous.

%% file: sections/02_history.tex
\section{A Brief History of Full-Duplex Spoken Dialogue Systems}
\label{sec:history}
\label{app:pre-fd}

\label{sec:history:proliferation}
\begin{figure*}[!htbp]
\centering
\includegraphics[width=\textwidth]{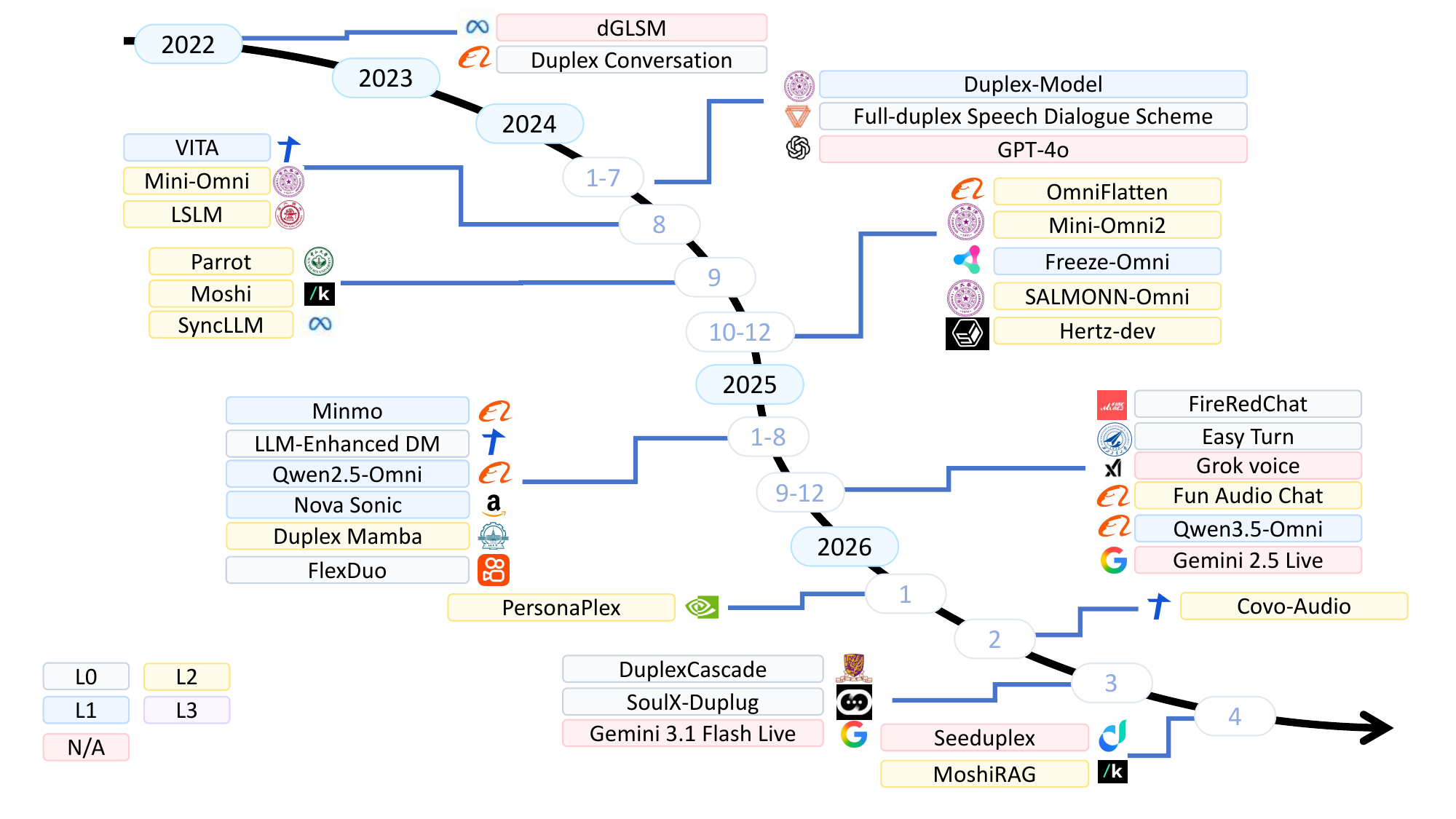}
\caption{Timeline of published full-duplex spoken dialogue systems, 2021--2026, grouped by the L0--L3 architectural layer defined below. N/A indicates that it is not applicable to full-duplex conversation classification or closed-source}
\label{fig:timeline}
\vspace{-8pt}
\end{figure*}

The history of FD-SDS is best read as three eras separated by two pivots. The pre-LLM era~\citep{sacks1974simplest,skantze2021turn} saw industrial L0 pipelines --- Google Duplex~\citep{leviathan2018google}, Ant Group~\citep{jin2021duplex}, and Alibaba DAMO~\citep{lin2022duplex} --- deploy VAD+EoT+DM stacks at production scale, with proprietary corpora that founded the open-source-versus-industrial data gap (\S\ref{sec:data}). The technical pivot came when speech became a token~\citep{zhang2023speechgpt,nguyen2023generative,borsos2023audiolm,wang2023neural,zhang2308speechtokenizer,defossez2022high}; the commercial pivot followed when GPT-4o (May 2024) made full-duplex a product expectation, with the immediate replication wave still half-duplex~\citep{xie2024mini,fang2025llama,zeng2024glm}. The third era opened with LSLM~\citep{ma2025language} and Moshi~\citep{defossez2024moshi} in late 2024; the L2 design space then fragmented across OmniFlatten~\citep{zhang2025omniflatten}, SyncLLM~\citep{veluri2024beyond}, and Mini-Omni2~\citep{xie2024miniomni2}, while MinMo~\citep{chen2025minmo} introduced the L1 hidden-state pattern, which subsequent FD industrial teams largely migrated past in favor of L2 token-level designs (Fun-Audio-Chat, Covo-Audio); the L1 shape itself nevertheless persisted in non-FD streaming designs (Qwen Thinker--Talker, Step-Audio R1.1). Late 2025 saw a modular comeback (FireRedChat~\citep{chen2025fireredchat}, FlexDuo~\citep{liao2025flexduo}, X-Talk~\citep{liu2025x}); 2026 brought further industrial parallel-stream L2 FD systems (Fun-Audio-Chat~\citep{team2025fun}, Covo-Audio~\citep{wang2026covo}), an L1-shaped non-FD streaming variant (Step-Audio~R1.1~\citep{wu2025mind}), plug-and-play L0 modules (SoulX-Duplug~\citep{yan2026soulx}, FastTurn~\citep{wang2026fastturn}), and L2 systems exploring non-standard backbones (DuplexMamba~\citep{lu2025duplexmamba}) and codec-free designs (SALMONN-omni~\citep{yu2026salmonn}). Adjacent streaming non-FD systems (GLM-4-Voice~\citep{zeng2024glm}, Qwen2.5-/3.5-Omni~\citep{xu2025qwen25omnitechnicalreport,team2026qwen3}, Step-Audio R1.1~\citep{wu2025mind}) share architectural shape but do not claim full-duplex behavior (\S\ref{sec:intro}, scope). Figure~\ref{fig:timeline} situates these on one timeline; the subsections below expand the pre-FD context and per-system narrative.

Stepping back, four architectural patterns recur across this proliferation and structure the rest of the paper. We refer to them as L0--L3 throughout and develop the cross-system audit in \S\ref{sec:hierarchy-audit}:
\vspace{0pt}
\begin{itemize}
  \setlength{\itemsep}{0pt}
  \item \textbf{L0 Module-level}: external scheduler outside the LLM (FireRedChat, FlexDuo, X-Talk, SoulX-Duplug).
  \item \textbf{L1 Hidden-state-level}: external module reads LLM hidden state (FD-claiming: MinMo, Freeze-Omni~\citep{wang2024freeze}; the Thinker--Talker of Qwen2.5/3.5-Omni and Step-Audio R1.1 share the same shape without claiming FD).
  \item \textbf{L2 Token-level}: duplex encoded in the token sequence (Moshi, LSLM, OmniFlatten, SyncLLM, Mini-Omni2, Fun-Audio-Chat, Covo-Audio, DuplexMamba on Mamba/SSM, SALMONN-omni codec-free; GLM-4-Voice shares the L2 architectural shape without claiming FD).
  \item \textbf{L3 Representation-level}: shared latent across streams; no published system yet.
\end{itemize}

\paragraph{Where prior surveys fall short.}
WavChat~\citep{ji2024wavchat} organizes systems along a cascaded/end-to-end trichotomy treating full-duplex as a binary attribute; Chen and Yu~\citep{chen2025turn} propose an Engineered-vs-Learned axis. Both collapse the design space onto a single axis. Three blind spots motivate this survey: i) neither distinguishes the stack layer at which the duplex decision is made (the L0--L3 convergence pattern of \S\ref{sec:hierarchy-audit}); ii) neither articulates \emph{which} interactions a full-duplex system should handle (\S\ref{sec:ontology}); iii) neither describes \emph{how} the system decides what to do at any instant (\S\ref{sec:state-machine}).

%% file: appendix/B_pre_fd_era.tex

\paragraph{Modular pipelines before the LLM era.}

Before ``full-duplex'' became a banner of the LLM era, turn-taking had been a research subject for half a century and a deployed product for several years. The cascaded speech-dialogue stack (ASR$\to$NLU$\to$DM$\to$NLG$\to$TTS) treated turn-taking as a scheduling decision driven by voice activity detection: when energy crossed a threshold and stayed silent for $\sim$300\,ms, the system declared the user's turn over and began responding. This worked acceptably for command-and-control interactions and badly for everything else.

The academic foundation traces back to Sacks, Schegloff, and Jefferson's 1974 \emph{Language} paper~\citep{sacks1974simplest}, which formalized turn-taking as a system of turn-constructional units and turn-allocation rules and remains the most-cited article in that journal. A modern computational lineage runs through Skantze and collaborators~\citep{skantze2021turn}, whose neural turn-taking-prediction work reframed speaker change as a continuous prediction problem rather than a VAD threshold and was the basis of much subsequent work on end-of-turn detection.

On the industrial side, the most visible pre-LLM full-duplex moment was \textbf{Google Duplex}, demonstrated by Sundar Pichai at Google I/O 2018, where an AI system phoned a hair salon and a restaurant and conducted multi-turn appointment-booking conversations in remarkably human-sounding speech~\citep{leviathan2018google}. Duplex was never published as an academic paper but it set the popular expectation that ``AI can carry on a phone conversation'' six years before GPT-4o, and its underlying L0 stack --- cascaded ASR + dialogue manager + TTS tuned for spontaneous timing --- defined the blueprint that industrial systems would refine in the years that followed. In Asia, Ant Group's 2021 Interspeech paper~\citep{jin2021duplex} described a three-module pipeline (speech-aware VAD, supervised end-of-turn classifier, response planner) running on real outbound calls, and Alibaba's DAMO Academy~\citep{lin2022duplex} followed in 2022 with a KDD paper that added speaker-conditioned VAD (pVAD) to filter background voices and reframed end-of-turn as a learned binomial decision over partial sentences, deployed in Alibaba's intelligent customer-service product. Together with Duplex, these systems established the L0 modular blueprint --- VAD plus EoT plus dialogue manager, each a separate model --- that LLM-era systems would later either inherit or argue against.

\paragraph{The end-to-end wave and the GPT-4o pivot.}

The first pivot was technical, not commercial: speech became a token. Discrete audio tokenizers~\citep{zhang2308speechtokenizer,defossez2022high} made it possible to feed waveforms into Transformer language models the same way text is fed, and a wave of end-to-end spoken dialogue models followed. SpeechGPT~\citep{zhang2023speechgpt} fused an LLM with speech encoding and decoding; AudioGPT~\citep{huang2024audiogpt} extended the multimodal recipe; Spirit-LM~\citep{nguyen2025spirit} demonstrated an interleaved speech-and-text language model; dGSLM~\citep{nguyen2023generative} --- chronologically earliest, in 2022 --- showed that two parallel speech streams could be jointly modelled without text at all, learning turn-taking rhythm and even laughter from the Fisher corpus. None of these were marketed as ``full-duplex'' in the modern sense. Adjacent lines of work from the same window --- audio-understanding LLMs such as Qwen2-Audio~\citep{chu2024qwen2} and integrated speech-interaction systems such as the broader Step-Audio line~\citep{huang2025step} --- share the token-level coupling between LLM and speech but target understanding or single-turn synthesis rather than duplex dialogue; we cite them here as historical context for the L1/L2 patterns that emerged around the same time.

The second pivot was commercial. On May 13, 2024, OpenAI's GPT-4o demonstration~\citep{gpt4o} showed conversational interactions in which users could interrupt the model mid-sentence and have it yield the floor smoothly. The wider community read this as proof that an end-to-end speech-to-speech model could be \emph{simultaneously} listening and speaking, and a replication race followed within months: Mini-Omni~\citep{xie2024mini}, LLaMA-Omni~\citep{fang2025llama}, and GLM-4-Voice~\citep{zeng2024glm} all shipped open implementations claiming end-to-end speech dialogue. Despite the end-to-end framing, a strict evaluation against the substantive-FD criterion shows these models largely functioned in a half-duplex manner: they could be \emph{interrupted} via a keyword trigger or a frontend VAD, but their decoder produced one stream at a time. Two bridge papers from this transitional window --- a full-duplex LLM scheme~\citep{wang2024full} and the Duplex-Model proposal from THUNLP~\citep{zhang2024beyond} --- attempted to paste a VAD-based scheduler onto an LLM without unifying the token layer; a token-level full-duplex system did not appear publicly until the next era.

\paragraph{The full-duplex proliferation.}

The third era opened with two papers one month apart. In August 2024, LSLM~\citep{ma2025language} demonstrated that a decoder-only TTS model could fuse a streaming SSL encoder in real time, with middle-layer fusion outperforming early or late variants. Kyutai publicly unveiled Moshi at a Paris launch event on July~3, 2024, and released both an interactive web demo and, in September 2024, the model weights and technical report~\citep{defossez2024moshi}. Moshi paired the Mimi neural audio codec (12.5\,Hz, 8 RVQ codebooks) with an RQ-Transformer that emitted three simultaneous streams (user-audio prediction, assistant audio, and an ``inner monologue'' text track), with roughly 200\,ms reported latency. Its training recipe is itself part of the historical record: pre-training on millions of hours of public audio, multi-stream post-training driven by speaker-diarized data, full-duplex fine-tuning on the Fisher corpus, and instruction tuning on a large body of TTS-synthesized dialogues~\citep{defossez2024moshi}. Together LSLM and Moshi reset what ``learned'' full-duplex was expected to mean.

Within six months, the design space had fragmented. OmniFlatten~\citep{zhang2025omniflatten} flattened four streams (user/assistant $\times$ speech/text) into a single sequence with a special silent-speech token; SyncLLM~\citep{veluri2024beyond} divided time into fixed chunks (160--240\,ms variants evaluated, with 160\,ms as the baseline) and alternated user and assistant token blocks with synchronization markers; Mini-Omni2~\citep{xie2024miniomni2} added a ``Stop Omni'' keyword interruption. GLM-4-Voice~\citep{zeng2024glm} shipped in the same window with a single-codebook tokenizer and a trillion-token pretraining mix; it shares the L2 architectural shape but does not explicitly claim full-duplex behavior.

Two further developments in 2025 are worth flagging because they recur in the framework discussion. First, MinMo~\citep{chen2025minmo} from Alibaba's FunAudioLLM team introduced the L1 pattern in the full-duplex setting: an LLM backbone whose hidden state is consumed by an external duplex-decision predictor. Notably, Qwen2.5-Omni~\citep{xu2025qwen25omnitechnicalreport} (and its successor Qwen3.5-Omni~\citep{team2026qwen3}) use a structurally identical pattern --- their Thinker--Talker decomposes generation by having the Talker read the Thinker's hidden state --- without claiming full-duplex behavior; they target streaming text+speech generation, where the ``two simultaneous outputs'' are text and speech modalities rather than overlapping user/assistant audio. The recurrence of the L1 architectural shape across these different product goals is what we treat as evidence that L1 is a structural attractor in \S\ref{sec:hierarchy-audit}, distinct from any specific FD claim. Second, late 2025 saw a modular comeback. FireRedChat~\citep{chen2025fireredchat} from Xiaohongshu and FlexDuo~\citep{liao2025flexduo} re-implemented the modular blueprint with stronger components; X-Talk~\citep{liu2025x} from SJTU's X-LANCE group went further and published an explicit advocacy paper arguing that the progression toward token-level designs is not the only viable trajectory, and that modular speech-to-speech systems deserve a serious long-term seat at the table. The era is not closed; it is contested.

Through late 2025 and 2026 the proliferation continued at industrial scale and along several non-mainstream directions. Fun-Audio-Chat~\citep{team2025fun} from the Tongyi Fun team adopted a parallel joint speech--text design in the spirit of Moshi, encoding the duplex decision at the token layer rather than via a hidden-state predictor; Covo-Audio~\citep{wang2026covo} appeared as a parallel L2 industrial system in early 2026, exposing the duplex transitions through THINK/SHIFT/BREAK tokens on the output stream; SoulX-Duplug~\citep{yan2026soulx} re-cast the modular blueprint as a plug-and-play streaming state-prediction module that can be attached to a frozen LLM; StepFun released Step-Audio~R1.1~\citep{wu2025mind}, a real-time streaming-generation variant built around a dual-brain articulation--reasoning split that does not claim full-duplex behavior but adopts the L1 hidden-state pattern, joining the Qwen Thinker--Talker lineage of L1-shaped non-FD systems. Two further L2 systems from this window broaden the L2 design space along axes that earlier L2 systems took as fixed: DuplexMamba~\citep{lu2025duplexmamba} replaces the Transformer decoder with a Mamba / state-space-model backbone, and SALMONN-omni~\citep{yu2026salmonn} drops the audio codec entirely and operates on continuous embeddings. Work like PersonaPlex~\citep{roy2026personaplex} (voice/role control) and MoshiRAG~\citep{chien2026moshirag} (asynchronous knowledge retrieval) extends dialogue capabilities beyond the core full-duplex setting and lies outside the scope of this survey.

%% file: sections/03_hierarchy_audit.tex
\section{Architectural Hierarchy and Cross-System Audit}
\label{sec:hierarchy-audit}

\begin{figure*}[!htbp]
\centering
\includegraphics[width=\textwidth, trim=0 4.5cm 0 0, clip]{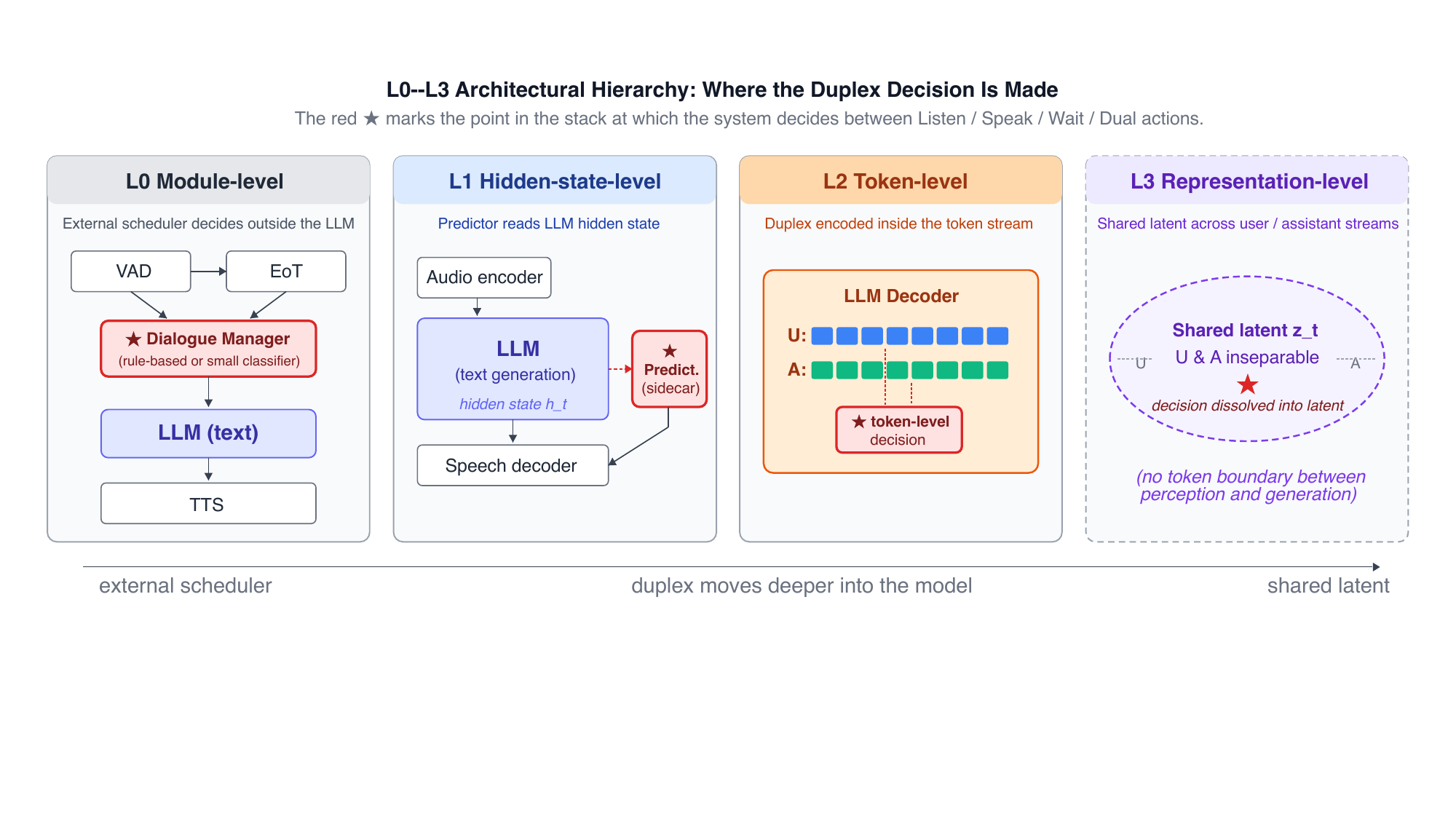}
\caption{The L0--L3 architectural hierarchy. The red \ding{72} marks the point in the stack at which the system decides between \textsc{Listen}, \textsc{Speak}, \textsc{Wait}, and \textsc{Dual} actions. The decision migrates from an external scheduler (L0) to a sidecar predictor reading the LLM hidden state (L1) to the token stream itself (L2) to a hypothetical shared latent (L3, not yet realized).}
\label{fig:l0-l3-arch}
\vspace{-8pt}
\end{figure*}

The L0--L3 hierarchy was introduced informally at the end of \S\ref{sec:history:proliferation} so that the historical sweep could refer to it without forward-referencing; this section makes it precise, visualizes it in Figure~\ref{fig:l0-l3-arch}, and uses it to audit representative systems in a single table. We treat the hierarchy as the first of the three frameworks because it directly organizes the systems already encountered: the T\,$\times$\,I\,$\times$\,R interaction ontology (\S\ref{sec:ontology}) and the decision state machine (\S\ref{sec:state-machine}) introduced in the next two sections then refine this view rather than precede it.

\subsection{Two Observations on the Hierarchy}
\label{sec:ha:observations}

Looking across the L0--L3 layers, two non-trivial patterns are visible. i) \textbf{L1 is a structural pattern recurring across product goals.} The L1 shape --- LLM hidden state read by an external module that produces the speech stream --- recurs in designs with quite different objectives: \citet{chen2025minmo} and \citet{wang2024freeze} use this pattern for full-duplex dialogue; \citet{xu2025qwen25omnitechnicalreport}'s Thinker--Talker and \citet{wu2025mind}'s Step-Audio R1.1 use the same shape for streaming text+speech generation without claiming FD. That independent teams converge on this pattern across different product goals suggests L1 is a structural attractor for LLM-based speech systems, not merely an artifact of one team's design choices. However, L1 is not the sole industrial path: an equally industrial slice of the field has migrated to L2 token-level designs (Fun-Audio-Chat, Covo-Audio, OmniFlatten), so the L1/L2 split now characterizes industrial FD systems. ii) \textbf{L0 remains contested rather than legacy.} \citet{liao2025flexduo}, \citet{chen2025fireredchat}, and the X-Talk position paper~\citep{liu2025x} are all 2025 arguments that the modular blueprint is competitive on latency, interpretability, and engineering cost, not merely a transitional stage before L2. The L0--L3 axis is therefore a description of design choices, not a ladder of progress.

\subsection{Cross-System Audit}
\label{sec:ha:audit}

Table~\ref{tab:audit} locates representative systems on the L-layer axis and summarizes which state-machine regions each layer can reach by construction.

Three patterns are visible in the table. i) The L2 row is the most populous, which is consistent with the academic narrative of the past two years, but L2 itself is heterogeneous: parallel-stream Moshi, flattened OmniFlatten, and chunk-alternated SyncLLM all reach \textsc{Dual} but with very different time granularities (continuous for parallel, frame-level for flattened, chunk-bounded for SyncLLM). ii) The L1 row contains the only \emph{convergent} pair in the table (MinMo and Qwen2.5-Omni Thinker--Talker arrived at structurally identical designs from different objectives), making L1 a confirmed structural attractor; production-grade FD deployments are nevertheless distributed across L1 and L2. iii) The L0 row is not empty and is not legacy: it is where new pluggable systems are still appearing in 2025--2026.

A subtler pattern is that architectural reachability is necessary but not sufficient. Two L2 systems can both reach \textsc{Dual} on paper yet behave very differently on the backchannel-during-speech cell, because reaching a state is a function of design and traversing it correctly is a function of training data. We return to this gap --- architecture sets capacity, training realizes it --- in \S\ref{sec:frontiers-conclusion} and in the evaluation appendix (\S\ref{app:evaluation}).

%% file: appendix/C_systems.tex

\subsection{L0 Module-Level Systems}
\label{app:systems}
\label{app:systems:l0}

L0 systems factor the duplex decision out of the generative model into one or more lightweight modules (typically VAD, end-of-turn (EoT) classifier, dialogue manager) that run continuously alongside an LLM-based or template-based generator. The LLM-era L0 systems inherit the topology established by the 2021 Interspeech and 2022 KDD industrial papers (\S\ref{app:pre-fd}); only the generator was replaced.

\paragraph{FireRedChat~\citep{chen2025fireredchat}.}
Three serial modules drive every transition. (i)~\textbf{pVAD}: a streaming personalized VAD with a speaker-condition channel, used to suppress background noise and competing speakers. (ii)~\textbf{EoT}: an end-of-turn classifier over streaming ASR partial results, reaching $\sim$96\% accuracy on Chinese and $\sim$95\% on English in the paper's evaluation. (iii)~\textbf{Dialogue Manager}: schedules the LLM/AudioLLM backend (FireRedChat supports both cascaded and semi-cascaded backends interchangeably). The pVAD covers the third-party-voice cell; the EoT covers the semantic-hesitation cell. Cooperative barge-in is handled acoustically and consequently fails on the backchannel-during-speech cell. Reported barge-in response time $T_{90} \approx 170$\,ms; reported end-to-first-response latency $P_{50} \approx 2.34$\,s, $P_{95} \approx 3.02$\,s (comparable to or better than the open frameworks LiveKit and Ten).

\paragraph{FlexDuo~\citep{liao2025flexduo}.}
A pluggable streaming module that decouples duplex control from the dialogue system. Notably introduces a third state called \textsc{Idle}, entered after the assistant finishes speaking, during which the assistant is silent and the system gates whether to remain idle or transition to \textsc{Listen}; the FlexDuo \textsc{Idle} therefore plays the role of an inter-turn buffer between \textsc{Speak} and \textsc{Listen} in our framework (see \S\ref{sec:state-machine}).

\paragraph{X-Talk~\citep{liu2025x}.}
A position paper from the SJTU X-LANCE group arguing that the L0 modular blueprint is competitive on latency, interpretability, and engineering cost, and is not merely a transitional stage before L2. Reframes the L0 trajectory as a parallel design path rather than a legacy.

\paragraph{SoulX-Duplug~\citep{yan2026soulx}.}
A plug-and-play streaming state-prediction module that attaches to a frozen LLM, restating the L0 modular blueprint with modern streaming components. Representative of the late-2025 modular renaissance.

\paragraph{Easy Turn~\citep{li2026easy}.}
Strictly a turn-taking detector rather than a full SDS: jointly fine-tuned on acoustic and linguistic modalities, predicting four states (complete / incomplete / backchannel / wait) that align directly with the standard-turn, hesitation, backchannel, and delayed-floor-yield cells of our ontology. Releases a 1145-hour open turn-taking corpus, currently the largest of its kind, and provides a drop-in replacement for legacy EoT modules.

\paragraph{FastTurn~\citep{wang2026fastturn}.}
A 2026 turn-detector module that, like Easy Turn, sits at the L0 layer and feeds a downstream LLM-based generator, but fuses acoustic and streaming semantic cues to make decisions \emph{before} the user's utterance is complete. Concretely, FastTurn pairs acoustic features (prosody, pitch, pauses, speech rhythm) with a streaming Connectionist Temporal Classification (CTC) decoder that emits partial transcripts in real time; the two signal paths are fused and consumed by a turn-decision head that distinguishes end-of-turn (standard-turn cell), hesitation, and backchannel cases. The design contrasts with Easy Turn's four-state classifier in two ways: (i)~by exposing partial-transcript semantics it can fire on the in-progress utterance rather than waiting for an end-of-turn token; (ii)~by combining the two modalities it remains robust under overlapping speech and noisy conditions where acoustic-only VAD or ASR-only EoT both degrade. The authors release a real-human-dialogue test set that captures backchannels, overlapping speech, environmental noise, and pitch variation, and report higher decision accuracy at lower interruption latency than representative baselines. FastTurn thus illustrates a continuing modular-EoT lineage that survives the L1/L2 token-level wave by competing on the L0-specific axes of low latency and deployment robustness, rather than abandoning them.

\paragraph{The L0 ceiling.} All L0 systems share a structural latency floor: even when the decision module fires in $\sim$100\,ms, the downstream LLM forward pass plus first-chunk TTS decoding pushes $\delta^{*}_{\text{respond}}$ to roughly 500\,ms. This is the physical motivation for the L1 step.

\subsection{L1 Hidden-State-Level Systems}
\label{app:systems:l1}

L1 systems retain modular separation but feed the decision module the LLM's hidden state $\mathbf{h}_t$, reusing the language model's semantic competence.

\paragraph{MinMo~\citep{chen2025minmo}.}
Voice Encoder $+$ Input Projector $\to$ Qwen LLM (LoRA), with the LLM's hidden state branching into (i) an Output Projector $\to$ Voice-Token LM $\to$ Token2Wav synthesis path, and (ii) a \emph{Full-Duplex Predictor} (single-layer Transformer with linear softmax, randomly initialized) that emits a binary decision in real time. The predictor operates in two modes: when the system is silent it discriminates ``respond now'' from ``keep waiting'' (standard-turn vs hesitation); when the system is speaking it discriminates barge-in from backchannel. Training reuses a 4000-h dialogue mix (real + simulated) under a multi-stage curriculum (speech-to-text $\to$ text-to-speech $\to$ speech-to-speech $\to$ duplex interaction alignment), with heuristic-rule auto-labeling for turn-taking events. The paper reports a speech-to-text latency of approximately 100\,ms and a full-duplex latency of approximately 600\,ms in theory, 800\,ms in practice.

\paragraph{Freeze-Omni~\citep{wang2024freeze}.}
Trains audio I/O around a \emph{frozen} LLM via a three-stage protocol (ASR, TTS, then a multi-task duplex stage on 60\,k multi-turn QA samples), using $\sim$8 GPUs total. Architecturally L1, but because the LLM weights are not updated by the duplex objective the model lives on the L0/L1 boundary --- a low-cost realization of the L1 pattern that trades ontology coverage for training simplicity.

\paragraph{Qwen2.5-Omni / Qwen3.5-Omni Thinker--Talker (non-FD, L1-shaped).}
Qwen2.5-Omni~\citep{xu2025qwen25omnitechnicalreport} and its successor Qwen3.5-Omni~\citep{team2026qwen3} are not marketed as full-duplex: their \emph{Thinker} (a text LLM) feeds its hidden representation to a \emph{Talker} (a dual-track autoregressive audio-token model) that streams text and speech outputs concurrently --- two output modalities, not overlapping user/assistant audio. The Thinker--Talker decomposition is nevertheless \emph{architecturally identical} to MinMo's LLM--Predictor pattern (external module reading LLM hidden state), which we treat in \S\ref{sec:hierarchy-audit} as evidence that the L1 hidden-state pattern recurs across product goals, beyond the FD niche alone.

\paragraph{Step-Audio R1.1 (non-FD, L1-shaped)~\citep{wu2025mind}.}
A 2026 real-time streaming-generation variant from the Step-Audio family with a dual-brain articulation--reasoning split. The paper does not claim full-duplex behavior; we include it as a taxonomy reference because its architecture --- an LLM-style ``Thinker'' whose hidden state is consumed by a downstream ``Articulator'' module --- recurs the L1 pattern of MinMo, providing additional cross-objective evidence that L1 is a structural attractor for LLM-based real-time speech systems.

\paragraph{L1 convergence.}
MinMo and Qwen2.5-Omni were proposed by different teams two months apart and arrived at structurally identical designs, despite targeting full-duplex dialogue and streaming text+speech generation respectively. Together with Freeze-Omni's frozen-LLM variant on the FD-claiming side and Step-Audio R1.1 on the non-FD streaming side, L1 remains a structural pattern shared across different product objectives. For industrial FD deployments specifically, L1 is now one of two parallel paths --- the other being L2 token-level design (Fun-Audio-Chat, Covo-Audio, OmniFlatten).

\subsection{L2 Token-Level Systems}
\label{app:systems:l2}

L2 systems eliminate the explicit decision module: turn-taking decisions are encoded in the token-level generative process itself. We organize L2 systems by their internal sub-mode (multi-stream, flatten, chunk-alternation, channel-fusion).

\paragraph{dGSLM (pre-history)~\citep{nguyen2023generative}.}
The first textless dual-channel speech LM: dual-tower Transformer with cross-attention, trained on the 2000-hour Fisher corpus with next-frame prediction, no text. Established that turn-taking rhythm and paralinguistic signals (laughter, sighs) can be learned end-to-end without text. Not a practical dialogue system (no instruction following) but the direct architectural ancestor of Moshi's multi-stream design.

\paragraph{LSLM~\citep{ma2025language}.}
Released August 2024, one month before Moshi --- the first published LLM-era dual-channel duplex system. Uses a token-based decoder-only TTS for the speaking channel and a streaming SSL encoder for the listening channel, fused at one Transformer layer (early/middle/late variants studied; middle layer optimal). The fusion-based duplex modelling is a distinct L2 sub-mode from Moshi's parallel-stream RQ-Transformer.

\paragraph{Moshi~\citep{defossez2024moshi}.}
Combines (i) the \textbf{Mimi} neural audio codec (12.5\,Hz, 8 RVQ codebooks; first codebook distilled from WavLM for semantic content, codebooks 2--8 for acoustic residuals), (ii) an \textbf{RQ-Transformer} (Temporal Transformer in the time dimension plus a Depth Transformer over the eight codebooks), and (iii) \textbf{Inner Monologue} (a text token stream temporally interleaved before the audio tokens, optionally with an acoustic delay). At every 80\,ms frame, the model simultaneously emits text tokens, a predicted user-audio token, and an assistant-audio token. Because the model predicts both streams at every frame, ``user begins speaking'' is naturally absorbed by the user-stream prediction and the assistant stream collapses to silence tokens without any explicit VAD/EoT module. The paper reports a theoretical latency of 160\,ms and approximately 200\,ms in practice. Training spans pre-training on roughly 7\,M hours of public audio with Whisper-transcript supervision, multi-stream post-training driven by speaker-diarized data, full-duplex fine-tuning on the Fisher corpus, and instruction tuning on synthetic dialogues generated by their own TTS.

\paragraph{OmniFlatten~\citep{zhang2025omniflatten}.}
Flattens four streams (user/assistant $\times$ speech/text) into a single sequence with a special \texttt{silent\_speech\_token}, then trains a standard GPT-style autoregressive model with no architectural modification, with text-stream tokens leading their corresponding speech-stream tokens. Demonstrates that full-duplex behavior does not require a bespoke decoder --- careful token design over a standard LM is sufficient. The paper reports low end-to-end latency without committing to a single number.

\paragraph{SyncLLM~\citep{veluri2024beyond}.}
Divides time into fixed-size chunks (the paper evaluates 160\,ms, 200\,ms, and 240\,ms variants; Figure~2 of the paper uses 160\,ms as the baseline) and alternates predicting the assistant's chunk and the user's chunk within a single sequence, marked by periodic synchronization tokens (\texttt{[S0]}/\texttt{[S1]}); uses deduplicated HuBERT tokens at 25\,Hz with interpolation. By the causal-influence criterion of \S\ref{sec:state-machine}, transitions occur only at chunk boundaries: within a chunk, the system cannot causally respond to the user's current audio. SyncLLM is therefore the canonical example of an L2-shaped but \emph{apparent} (not substantive) full-duplex system.

\paragraph{Mini-Omni / Mini-Omni2~\citep{xie2024mini,xie2024miniomni2}.}
Mini-Omni v1 generates text and audio tokens in parallel (same time step, two heads) on top of an arbitrary text LLM via the ``Any Model Can Talk'' adaptation recipe, trained on the VoiceAssistant-400K instruction corpus~\citep{xie2024mini}. Mini-Omni2 adds vision and a keyword-triggered interruption (``Stop Omni'') with state tokens \texttt{irq}/\texttt{n-irq} for command-mode barge-in detection. Outside the trigger keyword the model has no causal response to user audio --- archetypal apparent-FD behavior with strong open-source ecosystem value.

\paragraph{Fun-Audio-Chat~\citep{team2025fun}.}
A parallel joint speech--text model from the Tongyi Fun team. Following the multi-stream pattern of Moshi, it integrates explicit text and speech token streams that the LLM predicts jointly; the duplex decision is encoded at the token layer through stream coordination rather than read off the hidden state by a sidecar predictor, which places it in the L2 design space.

\paragraph{Covo-Audio~\citep{wang2026covo}.}
An industrial L2 system that exposes the duplex decision directly on the output token stream. The model continuously emits THINK tokens during listening, switches to its speaking turn by outputting a SHIFT token, and ends speech by emitting a BREAK token; transitions between \textsc{Listen}, \textsc{Speak}, and yield behaviors are therefore expressed at the token granularity rather than by an external module.

\paragraph{DuplexMamba~\citep{lu2025duplexmamba}.}
An L2 system that replaces the Transformer decoder with a Mamba / state-space-model backbone while keeping the duplex-in-token-sequence pattern. Demonstrates that L2 synchronization is backbone-agnostic and not tied to attention-based architectures.

\paragraph{SALMONN-omni~\citep{yu2026salmonn}.}
A standalone full-duplex speech LLM that operates on continuous embeddings rather than discrete audio-codec tokens. The duplex decision is still token-level autoregressive (continuous-valued token stream), keeping it within L2 by our definition; the contribution is to demonstrate that the L2 pattern does not strictly require a discrete codec tokenizer.

\paragraph{GLM-4-Voice (non-FD, L2-shaped)~\citep{zeng2024glm}.}
A streaming end-to-end speech chatbot with a single-codebook tokenizer at 175\,bps (12.5\,Hz, derived from an ASR model with a VQ bottleneck), pretrained on roughly a trillion tokens of mixed unsupervised, interleaved, and supervised data in Stage 1 and supervised fine-tuned in Stage 2. Architecturally adjacent to L2 with explicit turn structure preserved; the paper does not claim full-duplex behavior. Included here as evidence that the L2 architectural shape recurs even outside the FD-claiming niche.

\subsection{L3 Representation-Level: Why Empty}
\label{app:systems:l3}

L3 would require user and assistant streams to be encoded in a shared continuous latent rather than negotiated through discrete tokens. Three obstacles keep the L3 row empty in Table~\ref{tab:audit}: (i)~discrete audio tokenizers (Mimi~\citep{defossez2024moshi}, SpeechTokenizer~\citep{zhang2308speechtokenizer}, X-Codec~\citep{ye2025codec}, CosyVoice family~\citep{du2024cosyvoice}) have become the de-facto ecosystem, raising the cost of switching to continuous representations; (ii)~continuous-latent generative models (diffusion, flow matching) remain immature for streaming autoregressive use --- recent flow-matching detokenizers operate downstream of a discrete model, not at the modelling layer; (iii)~no agreed mathematical formulation of what L3 should look like exists yet. Three plausible directions are continuous audio latents with flow-matching modelling, JEPA-style dual-stream prediction, and world-model-conditioned dialogue. We discuss these in \S\ref{sec:frontiers-conclusion} as a concrete architectural agenda.

\subsection{Cross-System Comparison Tables}
\label{app:systems:tables}

Table~\ref{tab:audit} in the main text gives the L-layer / FSM-state matrix. Table~\ref{tab:ontology-coverage} (this section) gives the system $\times$ acid-test-cell coverage matrix for the seven cells most discussed in the paper.

\begin{table*}[t]
\centering
\small
\setlength{\tabcolsep}{3.5pt}
\renewcommand{\arraystretch}{1.25}
\begin{tabular}{l ccccccc}
\toprule
            & \textbf{Standard} & \textbf{Latched} & \textbf{Barge-in} & \textbf{Backchnl.} & \textbf{Third-party} & \textbf{Hesitation} & \textbf{Concurrent} \\
\textbf{System} & $T1 I1 R6$ & $T2 I1 R6$ & $T3 I4 R2$ & $T3 I2 R1$ & $T3 I7 R5$ & $T5 I6 R3$ & $T4 I1 R1$ \\
\midrule
FireRedChat   & \checkmark & $\vartriangle$ EoT     & \checkmark              & $\cdot$                 & \checkmark pVAD & \checkmark EoT       & $\cdot$ \\
MinMo         & \checkmark & $\vartriangle$ Pred.  & \checkmark              & \checkmark eval.\        & $\cdot$        & $\vartriangle$ partial & $\cdot$ \\
Freeze-Omni   & \checkmark & $\cdot$                & \checkmark              & $\cdot$                 & $\cdot$        & $\cdot$               & $\cdot$ \\
Moshi         & \checkmark & $\vartriangle$         & $\vartriangle$ supported & $\vartriangle$ training  & $\cdot$        & $\cdot$               & $\vartriangle$ arch. \\
LSLM          & \checkmark & $\cdot$                & \checkmark              & $\cdot$                 & $\cdot$        & $\cdot$               & $\cdot$ \\
OmniFlatten   & \checkmark & $\cdot$                & \checkmark              & \checkmark              & $\cdot$        & $\cdot$               & \checkmark \\
SyncLLM       & \checkmark & $\cdot$                & $\cdot$                 & \checkmark              & $\cdot$        & $\cdot$               & \checkmark \\
Mini-Omni2    & \checkmark & $\cdot$                & $\vartriangle$ keyword   & $\cdot$                 & $\cdot$        & $\cdot$               & $\cdot$ \\
\bottomrule
\end{tabular}
\caption{Ontology-cell coverage matrix for representative full-duplex systems, synthesized by combining each paper's own claims (abstract, methodology, evaluation) with our reading of evaluation evidence, training-data choices, and the ontology framework of this survey. \textbf{\checkmark}~the paper explicitly claims or evaluates the capability; \textbf{$\vartriangle$}~partial / qualified support (mentioned but not evaluated, partial mechanism, training-data only, command-mode keyword, or architectural plausibility without dedicated evaluation); \textbf{$\cdot$}~the paper does not claim the capability in any section we could verify. Cell shorthand: \textbf{Standard}~=~standard turn; \textbf{Latched}~=~latched zero-gap; \textbf{Barge-in}~=~cooperative barge-in; \textbf{Backchnl.}~=~backchannel during system speech; \textbf{Third-party}~=~third-party / noise rejection; \textbf{Hesitation}~=~hesitation with long silence; \textbf{Concurrent}~=~sustained concurrent speech. \emph{Absence of a claim or evidence does not imply that the system fails the cell --- only that, in our reading, the paper does not document the capability and we did not find independent evaluation evidence.} Three patterns are visible: (i)~no surveyed paper explicitly claims sustained concurrent speech, and only OmniFlatten and SyncLLM claim ``overlapping speech'' as an architecturally supported behavior; (ii)~third-party rejection is claimed only by FireRedChat via its pVAD module; (iii)~backchannel-during-speech is more widely claimed than we initially assumed --- MinMo reports back-channeling accuracy in the \textasciitilde70--80\% range; OmniFlatten and SyncLLM list backchanneling as supported behaviors; Moshi mentions backchannels only in its synthetic-training-data prompts.}
\label{tab:ontology-coverage}
\end{table*}

\subsection{Training-Recipe Cross-Comparison}
\label{app:systems:training}

A convergent four-stage training pattern emerged across L1/L2 systems through late 2024 and 2025: (1) general text LLM pre-training (often reused from an off-the-shelf backbone such as Qwen, LLaMA, or GLM); (2) speech-to-text alignment (ASR objective); (3) text-to-speech alignment (TTS objective); (4) duplex interaction alignment on full-duplex dialogue data, optionally followed by (4b) reinforcement learning with an interaction-quality reward. The data scarcity at stage 4 is the operative constraint: Moshi reports Fisher (2000\,h) plus large-scale synthesis; MinMo reports 4000\,h (3000\,h real, 1000\,h simulated); most other systems are dominated by synthesis. Auto-labeling at the duplex stage is universally adopted in some form --- MinMo's heuristic rules, OmniFlatten's silent-speech tokens, and SyncLLM's periodic anchors are three independent recipes, each enabling thousand-hour-scale training without manual turn-taking annotation. We treat this scarcity, and the strong dependence on auto-labeling, as the core of the data frontier discussed in \S\ref{sec:frontiers-conclusion} and \S\ref{app:data}.

%% file: sections/04_ontology.tex
\section{The T\,$\times$\,I\,$\times$\,R Interaction Ontology}
\label{sec:ontology}

\begin{table*}[t]
\centering
\small
\setlength{\tabcolsep}{6pt}
\renewcommand{\arraystretch}{1.35}
\begin{tabularx}{\textwidth}{@{}l@{\hskip 8pt} ccccc @{\hskip 10pt} >{\raggedright\arraybackslash}X@{}}
\toprule
\textbf{Layer}        & \textbf{\textsc{I}} & \textbf{\textsc{L}} & \textbf{\textsc{S}} & \textbf{\textsc{W}} & \textbf{\textsc{D}} & \textbf{Representative systems} \\
\midrule
\rowcolor{black!5}    \textbf{L0} Module    & \checkmark & \checkmark & \checkmark & $\vartriangle$ & $\cdot$        & FireRedChat, FlexDuo, X-Talk, SoulX-Duplug \\
\rowcolor{blue!7}     \textbf{L1} Hidden    & \checkmark & \checkmark & \checkmark & \checkmark     & $\vartriangle$ & MinMo, Freeze-Omni \\
\rowcolor{orange!10}  \textbf{L2} Token     & \checkmark & \checkmark & \checkmark & \checkmark     & \checkmark     & Moshi, LSLM, OmniFlatten, SyncLLM, Mini-Omni2, Fun-Audio-Chat, Covo-Audio, DuplexMamba (Mamba/SSM), SALMONN-omni (codec-free) \\
\rowcolor{violet!7}   \textbf{L3} Repr.\    & \multicolumn{5}{c}{\textit{--- not yet realized ---}}                                       & \textit{(open architectural frontier)} \\
\bottomrule
\end{tabularx}
\caption{Cross-system audit by L-layer with FSM-state reachability matrix. \textbf{\checkmark}~natively supported by construction; \textbf{$\vartriangle$}~limited or degenerate; \textbf{$\cdot$}~effectively unreachable. \textsc{I}/\textsc{L}/\textsc{S}/\textsc{W}/\textsc{D} = \textsc{Idle}/\textsc{Listen}/\textsc{Speak}/\textsc{Wait}/\textsc{Dual}. Row colours match the L0--L3 panels of Figure~\ref{fig:l0-l3-arch}. Only systems that explicitly claim full-duplex behavior are listed; Qwen2.5/3.5-Omni's Thinker--Talker and Step-Audio R1.1 (both L1-shaped) and GLM-4-Voice (L2-shaped) target streaming generation rather than FD (see \S\ref{sec:ha:observations}).}
\label{tab:audit}
\end{table*}

The L0--L3 hierarchy of \S\ref{sec:hierarchy-audit} organizes systems by \emph{where} the duplex decision is made; it does not yet say \emph{which} interactions the system should handle. We decompose a full-duplex interaction moment into a triple $(\text{T}, \text{I}, \text{R})$: the \emph{temporal relation} between user and system audio, the \emph{communicative intent} of the user's contribution, and the \emph{required system response}. Table~\ref{tab:tir-axes} lists all values on each axis. The three axes are deliberately orthogonal --- timing, meaning, and policy are distinct design questions, and conflating them is a common source of imprecision in current system descriptions.

\begin{table}[t]
\centering
\small
\setlength{\tabcolsep}{4pt}
\renewcommand{\arraystretch}{1.1}
\begin{tabular}{lll}
\toprule
\textbf{T (Temporal)} & \textbf{I (User Intent)} & \textbf{R (System Response)} \\
\midrule
T1 Sequential & I1 Information   & R1 Continue \\
T2 Latched    & I2 Backchannel   & R2 Stop \\
T3 Overlap    & I3 Repair        & R3 Wait \\
T4 Concurrent & I4 Floor-claim   & R4 Backchannel \\
T5 Silence    & I5 Floor-yield   & R5 Ignore \\
              & I6 Disfluency    & R6 Initiate \\
              & I7 Third-party   & \\
\bottomrule
\end{tabular}
\caption{The three orthogonal axes of the T\,$\times$\,I\,$\times$\,R interaction ontology. Every full-duplex interaction is a triple $(T_i, I_j, R_k)$.}
\label{tab:tir-axes}
\vspace{-8pt}
\end{table}

The temporal axis (T) ranges from clean sequential handoffs (T1) through latched, overlapping, and fully concurrent speech (T2--T4) to bilateral silence (T5)~\citep{sacks1974simplest,skantze2021turn}. The intent axis (I) extends standard speech-act vocabulary with the categories that matter for duplex modelling: backchannels (I2), floor-claims (I4) and floor-yields (I5), disfluencies (I6), and third-party audio (I7). The response axis (R) names the model's policy choice. The Cartesian product gives 210 nominal cells, most of which are physically impossible (e.g., R1 ``continue'' is meaningless when the system is not speaking) or pragmatically redundant, leaving a few dozen interaction types that a full-duplex system might face.

\paragraph{Six acid-test cells.} We single out six cells that exercise every axis and reveal current failure modes: \textbf{Standard turn}, the cascaded baseline; \textbf{Latched zero-gap}, which stresses end-of-turn latency; \textbf{Cooperative barge-in}, the scenario that GPT-4o's demo made famous; \textbf{Backchannel during system speech}, where most systems mis-classify the user's acknowledgement as a barge-in and stop --- the single most-cited failure mode in user feedback; \textbf{Third-party speech}, requiring speaker-conditioned VAD or robust speaker discrimination; and \textbf{Hesitation with long silence}, requiring semantic end-of-turn prediction. In our notation these are cells $(T_1, I_1, R_6)$, $(T_2, I_1, R_6)$, $(T_3, I_4, R_2)$, $(T_3, I_2, R_1)$, $(T_3, I_7, R_5)$, and $(T_5, I_6, R_3)$ respectively. Figure~\ref{fig:scenarios} (\S\ref{app:cells-detail}) concretizes each on User/Assistant audio timelines. To help readers form an intuitive understanding of the interaction logic of each of the six acid-test cells, we have also built a companion interactive demonstration website that plays back representative audio examples and visualizes the corresponding $(T, I, R)$ traces in real time.\footnote{\label{fn:demo}\url{https://duplexlm.github.io/DuplexLM/demo.html}}

The ontology does triple duty across the rest of the paper: as \textbf{data slices} (cell-coverage metric for training corpora, \S\ref{sec:data}), as \textbf{evaluation criteria} (the Full-Duplex-Bench series~\citep{lin2026full} aligns its overlap scenarios with specific cells, \S\ref{sec:evaluation}), and as \textbf{ablation targets} for the cross-system audit of \S\ref{sec:hierarchy-audit}. A practical consequence is that the ontology makes negative results legible: ``system X fails on the backchannel cell'' is a useful claim; ``system X is not always full-duplex'' is not. Prior efforts either cover human--human dialogue without the system-response axis~\citep{sacks1974simplest}, list capabilities without an axis structure~\citep{ji2024wavchat,chen2025turn}, or project onto the modular L0 setting only~\citep{li2026easy}.

%% file: appendix/C_cells.tex

\begin{figure*}[!htbp]
\centering
\includegraphics[width=\textwidth]{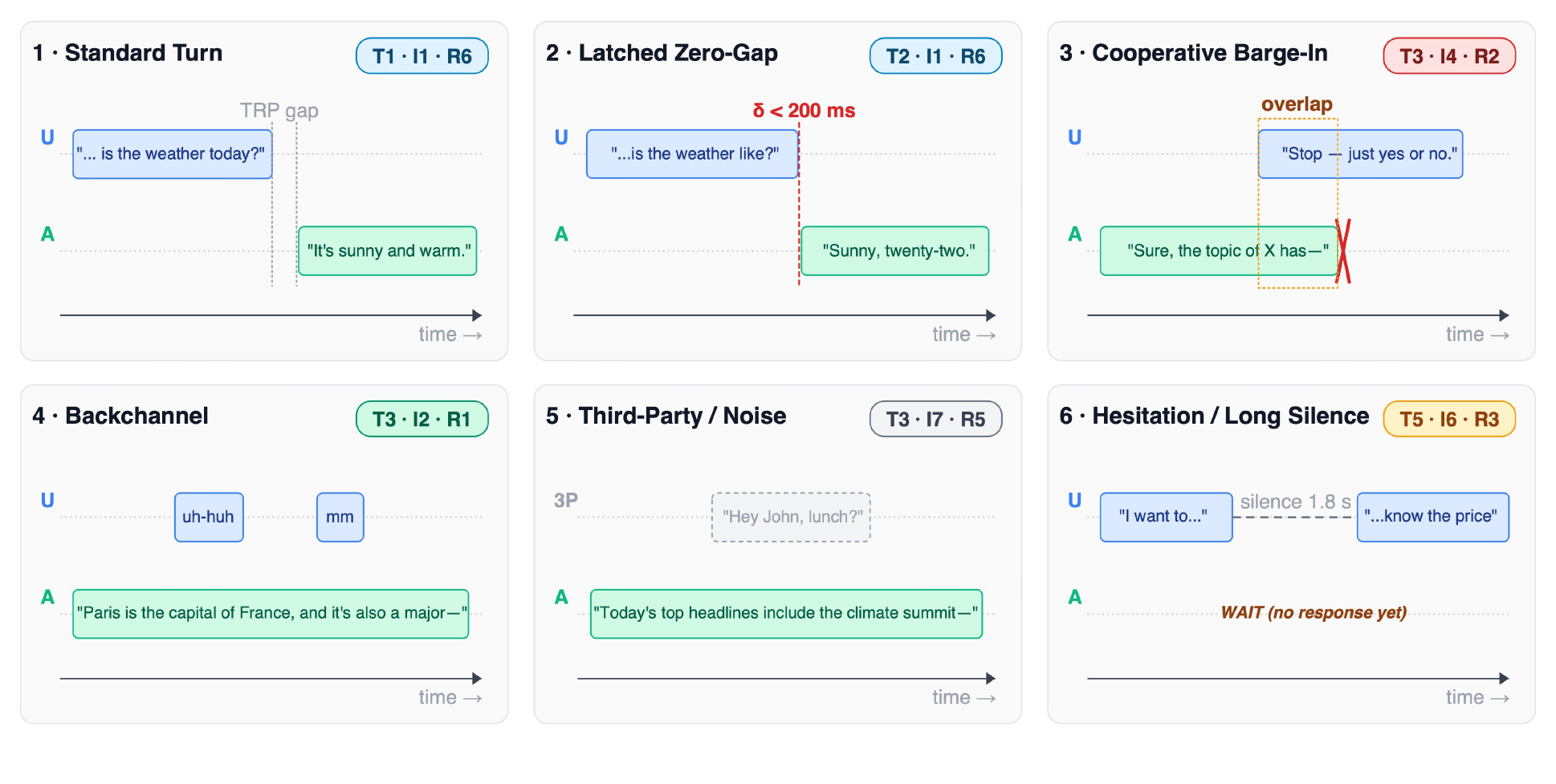}
\caption{Six canonical full-duplex interaction scenarios (referenced from \S\ref{sec:ontology}). Each panel concretizes one $(T, I, R)$ ontology cell on User (U) / Assistant (A) audio timelines. Interactive audio examples and real-time visualization of the corresponding traces are available at our companion demo site\protect\footref{fn:demo}.}
\label{fig:scenarios}
\vspace{-8pt}
\end{figure*}

\label{app:cells-detail}
\subsection{Standard Turn $(T1, I1, R6)$}

\textbf{Behavior.} Clean handoff at a transition-relevance place (TRP), with a 200--600\,ms inter-turn gap.\\
\textbf{Coverage.} Every system since cascaded SDS handles this; it is the baseline of any duplex evaluation.

\subsection{Latched Zero-Gap $(T2, I1, R6)$}

\textbf{Behavior.} Near-zero handoff. The assistant must already have begun generating before the user finishes (streaming generation + early end-of-turn prediction).\\
\textbf{Coverage.} Requires semantic VAD plus speculative TTS. Moshi, MinMo, and FireRedChat reach this; cascaded baselines miss because their EOT decision happens after silence has been observed.

\subsection{Cooperative Barge-In $(T3, I4, R2)$}

\textbf{Behavior.} The user asserts the floor while the assistant is still talking. The assistant must stop within roughly 200\,ms for the interaction to feel substantively full-duplex.\\
\textbf{Pitfall.} Apparent-FD systems (chunk-aligned designs, keyword-triggered interruption in command-mode systems) keep generating across the chunk boundary --- they fail the causal-influence criterion (\S\ref{app:evaluation}).

\subsection{Backchannel During System Speech $(T3, I2, R1)$}

\textbf{Behavior.} Short overlapping signals from the user (``uh-huh'', ``mm'', ``\textit{dui}'') are listener support, not a floor claim --- the assistant must keep going, possibly with subtle prosodic accommodation.\\
\textbf{Pitfall.} Na\"ive acoustic VAD treats any user voice as barge-in, fragmenting the system mid-clause. Distinguishing $I2$ from $I4$ requires intent classification, not just energy detection.

\subsection{Third-Party / Noise $(T3, I7, R5)$}

\textbf{Behavior.} Speech that overlaps the assistant but does not come from the addressed user (third party, TV, environmental speech) must be filtered out and ignored.\\
\textbf{Mechanism.} Personal VAD (pVAD), speaker embeddings, or an LLM-level relevance gate (sometimes called ``semantic denoise''). FireRedChat's pVAD module and the Ant Group / Alibaba L0 pipelines target precisely this cell.

\subsection{Hesitation with Long Silence $(T5, I6, R3)$}

\textbf{Behavior.} Acoustic silence does not imply end-of-turn. Semantic content ``I want to\ldots'' is incomplete; the system should wait rather than treat the silence as a turn-yielding cue.\\
\textbf{Pitfall.} Fixed-threshold VAD (e.g., $\sim$500\,ms) fires prematurely. Semantic end-of-turn prediction~\citep{skantze2021turn} or an LLM-conditioned predictor is required.

%% file: sections/05_state_machine.tex
\section{The Full-Duplex Decision State Machine}
\label{sec:state-machine}
\label{app:state-machine}

The hierarchy of \S\ref{sec:hierarchy-audit} names \emph{where} the duplex decision is made and the ontology of \S\ref{sec:ontology} names \emph{which} interactions the system faces. The decision state machine names what the system is \emph{doing} at each instant. Earlier FSM formulations are tightly tied to a particular system: a two-state \textsc{Speak}/\textsc{Listen} machine~\citep{wang2024full} and FlexDuo's three-state extension~\citep{liao2025flexduo}.\footnote{FlexDuo's \textsc{Idle} is an inter-turn silent buffer entered after the assistant finishes speaking, during which the system decides whether to remain idle or transition to \textsc{Listen}; functionally it sits between our \textsc{Speak} and \textsc{Listen} states rather than coinciding with our classical \textsc{Idle}.} We extend that lineage along two axes: we add a \textsc{Wait} state (semantic hesitation handling) and a \textsc{Dual} state (overlap-aware integration), and we couple every state and transition to a specific $(T, I, R)$ ontology cell so that the FSM works as a cross-system survey instrument.

\subsection{Five States, Eleven Transitions}
\label{sec:sm:states}

We model a full-duplex system as occupying, at every instant, one of five states:

\begin{itemize}
  \setlength{\itemsep}{1pt}
  \item \textsc{Idle} --- neither user nor system is producing speech.
  \item \textsc{Listen} --- only the user is speaking; the system is silently encoding.
  \item \textsc{Speak} --- only the system is speaking; user audio is silent or below a noise floor.
  \item \textsc{Wait} --- the system has paused inside its own turn (after a hesitation or unfinished user utterance) and is monitoring before resuming.
  \item \textsc{Dual} --- both speakers are producing audio with active mutual integration (a sustained backchannel during system speech, a third-party voice the system is filtering out, or true concurrent speech).
\end{itemize}

\begin{figure*}[!htbp]
\centering
\includegraphics[width=0.95\textwidth, trim=0 2cm 0 0, clip]{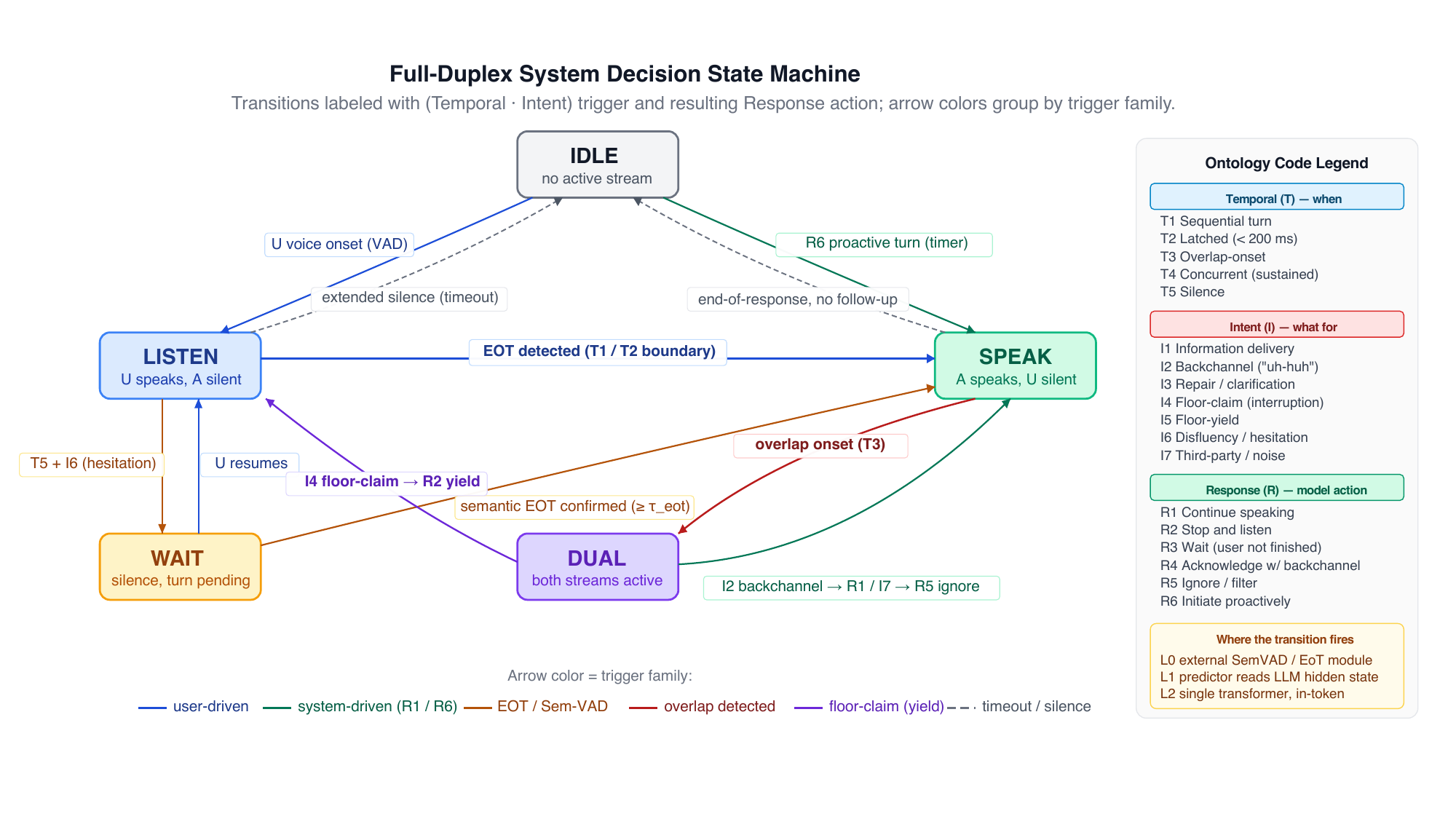}
\caption{The full-duplex decision state machine: five states and eleven transitions. Each transition is labelled by its $(T, I)$ trigger and the resulting $R$ action; arrow colors group transitions by trigger family.}
\label{fig:state-machine}
\vspace{-8pt}
\end{figure*}

Eleven transitions connect these five states; the full transition graph is given in Figure~\ref{fig:state-machine}. Each transition is labelled with the $(T, I)$ trigger that causes it and the resulting $R$ action. The transitions split naturally into three families. i) \emph{Onset transitions} (\textsc{Idle}$\to$\textsc{Listen}, \textsc{Idle}$\to$\textsc{Speak}) are driven by VAD or by a proactive-initiation timer (R6). ii) \emph{Turn-handoff transitions} (\textsc{Listen}$\to$\textsc{Speak}, \textsc{Speak}$\to$\textsc{Idle}, \textsc{Wait}$\to$\textsc{Speak}) are gated by end-of-turn prediction in the T1 / T2 regime and by semantic-EOT confirmation when the user's turn includes I6 disfluency. iii) \emph{Overlap transitions} (\textsc{Speak}$\to$\textsc{Dual}, \textsc{Dual}$\to$\textsc{Speak}, \textsc{Dual}$\to$\textsc{Listen}) are the duplex-specific transitions that distinguish a true full-duplex system from a half-duplex one; the destination depends on the intent $I$ of the overlapping audio (I2 backchannel stays in \textsc{Dual} or returns to \textsc{Speak}; I4 floor-claim transitions to \textsc{Listen}; I7 third-party noise is ignored and the system returns to \textsc{Speak}).

\paragraph{Composing with the ontology.} Each ontology cell corresponds to a small set of transitions: cooperative barge-in is \textsc{Speak}$\to$\textsc{Dual}$\to$\textsc{Listen}; backchannel-during-speech is a brief \textsc{Speak}$\to$\textsc{Dual}$\to$\textsc{Speak} excursion; third-party speech is the same shape with a different intent classification; hesitation is \textsc{Listen}$\to$\textsc{Wait}$\to$\textsc{Listen}. The state machine therefore turns ``what does system $X$ do in cell $Y$?'' into a structured lookup, and surfaces a per-system constraint the ontology alone cannot: which transitions a given L-layer can architecturally attempt.

%% file: appendix/F_state_machine_proofs.tex

\subsection{Notation}
\label{app:sm:notation}

A state is one of $\mathcal{S} = \{\textsc{Idle}, \textsc{Listen}, \textsc{Speak}, \textsc{Wait}, \textsc{Dual}\}$. A transition is written $s \xrightarrow{(T_i, I_j) / R_k} s'$ to mean: when the system is in state $s$ and the current input is of temporal class $T_i$ and intent class $I_j$, the system takes the response of class $R_k$ and moves to state $s'$. Triggers and actions are drawn from the T\,$\times$\,I\,$\times$\,R ontology of \S\ref{sec:ontology}. A transition with no $R$ component is a perceptual transition (the system updates its state based on input without producing output). Self-loops are written $s \xrightarrow{(T_i, I_j) / R_k} s$.

\subsection{The Eleven Transitions}
\label{app:sm:transitions}

\begin{table*}[t]
\centering
\small
\setlength{\tabcolsep}{4pt}
\renewcommand{\arraystretch}{1.2}
\begin{tabular}{l l l l l p{3.8cm}}
\toprule
\textbf{\#} & \textbf{From} & \textbf{To} & \textbf{Trigger $(T, I)$} & \textbf{Action $R$} & \textbf{Notes} \\
\midrule
$\tau_1$  & \textsc{Idle}   & \textsc{Listen} & $(T_1, I_1)$ user onset             & ---            & Front-end VAD or speaker-conditioned VAD detects user speech. \\
$\tau_2$  & \textsc{Idle}   & \textsc{Speak}  & $(T_1, I_5)$ proactive timer        & $R_6$ respond  & System initiates a turn after a silence threshold. \\
$\tau_3$  & \textsc{Listen} & \textsc{Speak}  & $(T_1, I_1)$ EoT confirmed          & $R_6$ respond  & End-of-turn predictor fires; standard handoff. \\
$\tau_4$  & \textsc{Listen} & \textsc{Wait}   & $(T_5, I_6)$ user hesitation        & $R_3$ wait     & Semantic EoT says the user has not finished. \\
$\tau_5$  & \textsc{Wait}   & \textsc{Speak}  & $(T_1, I_1)$ EoT re-confirmed       & $R_6$ respond  & User completes their utterance. \\
$\tau_6$  & \textsc{Wait}   & \textsc{Listen} & $(T_3, I_1)$ user resumes           & ---            & User keeps talking after the pause. \\
$\tau_7$  & \textsc{Speak}  & \textsc{Idle}   & $(T_1, I_1)$ system EoT             & ---            & System finishes its turn cleanly. \\
$\tau_8$  & \textsc{Speak}  & \textsc{Dual}   & $(T_3, I_*)$ overlap detected       & ---            & User audio onset while system is speaking; $I$ disambiguated in \textsc{Dual}. \\
$\tau_9$  & \textsc{Dual}   & \textsc{Listen} & $(T_3, I_4)$ floor-claim            & $R_2$ yield    & System stops speaking and listens. \\
$\tau_{10}$ & \textsc{Dual} & \textsc{Speak}  & $(T_3, I_2)$ backchannel           & $R_1$ continue & System keeps speaking; brief listener-acknowledgement absorbed. \\
$\tau_{11}$ & \textsc{Dual} & \textsc{Speak}  & $(T_3, I_7)$ third-party / noise   & $R_5$ ignore   & System keeps speaking; filters out non-addressee audio. \\
\bottomrule
\end{tabular}
\caption{The eleven transitions of the full-duplex decision state machine. Triggers are written as $(T, I)$ pairs from the ontology of \S\ref{sec:ontology}; actions are $R$ classes. Transitions $\tau_1$, $\tau_6$, $\tau_7$, and $\tau_8$ are perceptual --- they update state from input without producing an output action. Sustained $T_4$ concurrent speech is the \textsc{Dual} self-loop with no transition out, omitted from the row count because every system in \S\ref{app:systems} fails to enter it stably.}
\label{tab:transitions}
\end{table*}

Table~\ref{tab:transitions} enumerates the eleven transitions. The transitions split into three families introduced in \S\ref{sec:sm:states}: \emph{onset transitions} ($\tau_1$, $\tau_2$) bring the system out of \textsc{Idle}; \emph{turn-handoff transitions} ($\tau_3$--$\tau_7$) move between single-speaker states; \emph{overlap transitions} ($\tau_8$--$\tau_{11}$) describe what happens when both speakers are active. The overlap family is the duplex-specific one: a half-duplex system implements $\tau_8$ as an unconditional fast handoff to \textsc{Listen} and does not distinguish $\tau_9$ from $\tau_{10}$ or $\tau_{11}$, which is the formal sense in which it ``cannot tell backchannel from interruption.''

\subsection{Cell-to-Transition Mapping}
\label{app:sm:mapping}

The state machine and the T\,$\times$\,I\,$\times$\,R ontology compose: each non-trivial ontology cell maps to exactly one transition, and each transition realizes a small set of cells.

\begin{itemize}
  \setlength{\itemsep}{1pt}
  \item $T_1 \cdot I_1 \cdot R_6$ \textit{Standard} = $\tau_3$.
  \item $T_2 \cdot I_1 \cdot R_6$ \textit{Latched} = $\tau_3$ executed with zero or negative latency between user EoT and system speech onset.
  \item $T_3 \cdot I_4 \cdot R_2$ \textit{Cooperative barge-in} = $\tau_8$ followed by $\tau_9$.
  \item $T_3 \cdot I_2 \cdot R_1$ \textit{Backchannel} = $\tau_8$ followed by $\tau_{10}$.
  \item $T_3 \cdot I_7 \cdot R_5$ \textit{Third-party} = $\tau_8$ followed by $\tau_{11}$.
  \item $T_5 \cdot I_6 \cdot R_3$ \textit{Hesitation} = $\tau_4$ followed by $\tau_5$ (or $\tau_6$).
  \item $T_4 \cdot I_1 \cdot R_1$ \textit{Concurrent} = $\tau_8$ followed by a sustained \textsc{Dual} self-loop.
\end{itemize}

A system that lacks one of the eleven transitions also lacks every cell that depends on it. This is the precise sense in which the state machine functions as an ablation inventory: each transition is an independently testable behavior, and a benchmark score that fails on one cell can be localized to the missing or incorrectly implemented transition.

\subsection{Per-System Traces}
\label{app:sm:traces}

We work through six traces against the table above to illustrate how systems at different L-layers realize the same FSM.

\paragraph{Trace 1 --- Standard turn-taking (every system).} $\textsc{Idle} \xrightarrow{\tau_1} \textsc{Listen} \xrightarrow{\tau_3} \textsc{Speak} \xrightarrow{\tau_7} \textsc{Idle}$. This is the half-duplex baseline trace. Every system in \S\ref{app:systems} traverses it; differences live in the timing of $\tau_3$ and $\tau_7$, not in the trace itself.

\paragraph{Trace 2 --- Hesitation handling (MinMo).} $\textsc{Listen} \xrightarrow{\tau_4} \textsc{Wait} \xrightarrow{\tau_5} \textsc{Speak}$. MinMo~\citep{chen2025minmo} fires $\tau_4$ when its semantic EoT predictor recognizes that the user has not yet finished (e.g., trailing-off intonation or an incomplete syntactic frame). FireRedChat~\citep{chen2025fireredchat} executes the same trace at L0 via its EoT classifier; the difference is implementation layer, not the trace shape. An L2 system without an explicit semantic-EoT module typically collapses $\tau_4$ into a delayed $\tau_3$ --- which is observable as a longer mean response latency on $T_5$-heavy test cases but not as a separate transition.

\paragraph{Trace 3 --- Cooperative barge-in (Moshi).} $\textsc{Speak} \xrightarrow{\tau_8} \textsc{Dual} \xrightarrow{\tau_9} \textsc{Listen}$. Moshi~\citep{defossez2024moshi} executes $\tau_8$ implicitly because the user-audio stream is always being predicted in parallel with assistant audio; the system is therefore in \textsc{Dual} continuously and $\tau_9$ amounts to suppressing the assistant audio stream while continuing to attend to the user stream. The transition latency is dominated by the codec frame rate (12.5\,Hz $\to$ 80\,ms minimum) plus a small recognition delay.

\paragraph{Trace 4 --- Backchannel-during-speech (OmniFlatten).} $\textsc{Speak} \xrightarrow{\tau_8} \textsc{Dual} \xrightarrow{\tau_{10}} \textsc{Speak}$. OmniFlatten~\citep{zhang2025omniflatten} uses an explicit \texttt{silent\_speech\_token} on the assistant stream to indicate ``stay silent here'' and continues the assistant utterance after the user's brief acknowledgement. The trace is short --- typically a few tokens at the flattened sequence's granularity. A system that lacks $\tau_{10}$ collapses this trace into $\tau_8 + \tau_9$ (backchannel mistaken for floor-claim, system yields incorrectly).

\paragraph{Trace 5 --- Third-party voice (FireRedChat).} $\textsc{Speak} \xrightarrow{\tau_8} \textsc{Dual} \xrightarrow{\tau_{11}} \textsc{Speak}$. FireRedChat's~\citep{chen2025fireredchat} speaker-conditioned VAD (pVAD) detects that the overlapping audio is not from the addressed user; the dialogue manager routes this to $\tau_{11}$ rather than $\tau_9$. The decision is made at the module layer rather than the token layer, but the FSM trace is identical to Moshi's structurally; the difference is which component is responsible for the $I_7$ classification.

\paragraph{Trace 6 --- Sustained concurrent (no system).} $\textsc{Speak} \xrightarrow{\tau_8} \textsc{Dual} \xrightarrow{\text{(self-loop)}} \textsc{Dual} \cdots$. The state machine admits a \textsc{Dual} self-loop on $T_4 \cdot I_1 \cdot R_1$, but no system in \S\ref{app:systems} stably sustains it. The architectural cause varies (chunk-alternating designs cannot represent it; multi-stream designs can represent it but fail to learn it), but in every case the trace collapses to a $\tau_8 + \tau_9$ or to acoustic degradation within a few frames. Sustained $T_4$ is the open transition of the state machine.

\subsection{What the State Machine Is Not}
\label{app:sm:scope}

We close with three explicit non-claims, so that the formalism is not over-read.

\emph{First}, the state machine is not a controller. It describes the externally observable behavior of a full-duplex system; nothing in it specifies how a particular system computes its next transition. An L0 system computes it from module outputs, an L1 system from a sidecar predictor, an L2 system from token-level decoding --- and the FSM is silent on which.

\emph{Second}, the transitions are not equally hard. $\tau_3$ and $\tau_7$ (clean turn-taking) are essentially solved by every system in the audit; $\tau_{10}$, $\tau_{11}$, and the sustained \textsc{Dual} self-loop on $T_4$ are not. A flat per-transition checklist would obscure this; the cell-coverage matrix of \S\ref{app:systems} is the appropriate companion view.

\emph{Third}, the eleven transitions are not exhaustive over all conceivable dialogue events --- they cover the cells the ontology of \S\ref{sec:ontology} identifies as currently load-bearing. A future ontology revision that introduced, for instance, an $I_8$ (sustained meta-conversational signal such as long-form laughter) would add a corresponding transition; the framework is open in this direction.

%% file: sections/06_data.tex
\section{The Data Bottleneck}
\label{sec:data}
\label{app:data}

The audit of \S\ref{sec:hierarchy-audit} shows that architectural reachability is necessary but not sufficient: two L2 systems can both reach \textsc{Dual} on paper yet behave very differently on the backchannel-during-speech cell. We argue that across the L0--L2 systems audited, the binding constraint on system-level full-duplex behavior is not the architectural choice but the cell-coverage distribution of training data. Three pieces of evidence motivate this thesis.

\emph{First}, within the L2 row of Table~\ref{tab:audit}, the five token-level FD systems (Moshi, LSLM, OmniFlatten, SyncLLM, Mini-Omni2) are architecturally eligible to reach the same cells, but their actual cell coverage (Table~\ref{tab:ontology-coverage}, \S\ref{app:systems}) varies substantially without aligning to sub-design choices: OmniFlatten and SyncLLM use different sub-designs but both claim backchannel-during-speech and sustained concurrent coverage; Mini-Omni2 shares OmniFlatten's flattened family but claims neither. The aligning factor is what data each system was post-trained on. \emph{Second}, Moshi's paper documents that its full-duplex fine-tuning stage uses the Fisher corpus~\citep{cieri2004fisher,defossez2024moshi} as a primary source of real multi-stream conversational audio, and identifies this stage as where its turn-taking competence is acquired (without reporting an explicit remove-Fisher ablation). \emph{Third}, MinMo's L1 predictor is a one-layer Transformer that reaches usable barge-in / backchannel behavior largely because of the 4000-hour dialogue mix (3000\,h real, 1000\,h simulated) that the team annotated with heuristic turn-taking labels~\citep{chen2025minmo}; comparable predictors without the same data scale do not.
\vspace{-2pt}

\paragraph{Two-track structural asymmetry.} Public training data ($\sim$5\,k\,h total) is dominated by Fisher, Switchboard~\citep{godfrey1992switchboard}, and the CallHome multilingual telephone series\footnote{LDC catalog: CallHome American English Speech (LDC97S42), CallHome Mandarin Chinese (LDC96S34), CallHome Japanese (LDC96S37), CallHome Spanish (LDC96S35), CallHome German (LDC97S43), and CallHome Egyptian Arabic (LDC97S45). Distributed by the Linguistic Data Consortium.}; recent additions are CANDOR~\citep{reece2023candor} (Zoom calls), MagicData-RAMC~\citep{yang2022magicdata} (Mandarin mobile), Easy Turn~\citep{li2026easy} (turn-class labels), and HumDial Track\,2~\citep{wang2026humdial} (a public corpus with explicit third-party-speech sub-tasks, albeit with a small training subset for that cell). Industrial proprietary data is generally undisclosed in scale and in cell coverage; what is publicly known suggests deploying-product-specific bias (e.g., individual call-center reports indicate over-representation of certain overlap patterns), but a uniform comparison against Track-A is not currently possible from public sources. The asymmetry directly explains the open-source-versus-industrial gap in published FD behavior; \S\ref{app:data} gives the full corpus inventory and the cell-by-cell coverage matrix.
\vspace{0pt}
\paragraph{What the data audit predicts.} The cells systematically under-served in public training data --- sustained concurrent speech ($T_4$) and, until 2026, third-party voice during system speech --- are the same cells on which every audited system fails to demonstrate robust performance in current evaluations. The forthcoming HumDial Track\,2 release closes the third-party gap by design but leaves sustained concurrent speech open; we treat $T_4$ concurrent speech as the primary residual data frontier and identify five concrete data directions (synthesis pipelines, third-party injection, cross-dataset alignment, real concurrent-speech collection, RL-with-conversational-rewards) in \S\ref{app:data:agenda}.

The thesis is bounded: it applies at L0--L2 (the L3 frontier is an architecture problem, not a data problem; see \S\ref{sec:frontiers-conclusion}); it claims that data sets the \emph{ceiling} of behavior while architecture sets the gain; and it is conditional on the current evaluation regime, which the next section discusses.

%% file: appendix/D_data.tex

\subsection{Three Training-Data Types}
\label{app:data:types}

We classify training data for full-duplex systems into three types by the completeness of dialogue state they carry:

\begin{itemize}
  \setlength{\itemsep}{1pt}
  \item \textbf{Type A --- Single-stream speech}: single-channel audio without dialogue structure (ASR-style corpora). Used at the speech-text alignment stage; supplies the LLM's speech-token vocabulary but does not teach duplex dynamics.
  \item \textbf{Type B --- Single-stream dialogue / instruction}: single-channel question-answer or instruction data (VoiceAssistant-400K~\citep{xie2024mini}, InstructS2S-200K~\citep{fang2025llama}). Used to teach ``answer in speech'' but not duplex timing.
  \item \textbf{Type C --- Two-stream time-synchronous dialogue}: two-channel recordings (user and assistant on separate channels) with frame-level time alignment. Fisher, Switchboard, CANDOR, Easy Turn, internal call-center corpora. \textbf{This is the only data type that directly teaches full-duplex behavior.}
\end{itemize}

Type C is the binding constraint. Real-world collection requires two human speakers, dual-channel hardware, frame-level timestamping, content annotation, and privacy compliance; per-hour cost is order-of-magnitude higher than single-stream ASR data. The remainder of this section focuses on Type C.

\subsection{Public Type-C Corpora}
\label{app:data:corpora}

\begin{table*}[t]
\centering
\small
\setlength{\tabcolsep}{6pt}
\renewcommand{\arraystretch}{1.2}
\begin{tabular}{l r l l l}
\toprule
\textbf{Corpus} & \textbf{Hours} & \textbf{Era} & \textbf{Lang.} & \textbf{Format} \\
\midrule
\multicolumn{5}{l}{\textit{Two-speaker telephone (LDC era)}} \\
Switchboard      & $\sim$260\,h  & 1992    & EN          & 2-channel \\
CallHome         & $\sim$60\,h (EN) & 1990s   & EN/ZH/JP/AR/DE/ES & 2-channel \\
Fisher           & $\sim$2000\,h & 2003--04 & EN         & 2-channel \\
\midrule
\multicolumn{5}{l}{\textit{Multi-party meeting}} \\
ICSI Meeting     & $\sim$72\,h   & 2003    & EN          & multi-channel \\
AMI Meeting      & $\sim$100\,h  & 2007    & EN          & multi-channel + video \\
\midrule
\multicolumn{5}{l}{\textit{Modern conversational (video / mobile)}} \\
MagicData-RAMC   & 180\,h        & 2022    & ZH          & mono (smartphone) \\
CANDOR           & $\sim$850\,h  & 2023    & EN          & 2-channel (Zoom) \\
\midrule
\multicolumn{5}{l}{\textit{Turn-taking labeled / FD-by-design}} \\
Easy Turn train  & 1145\,h       & 2025    & EN+ZH       & turn-taking detection \\
HumDial Track\,2 & 107\,h        & 2026    & EN+ZH       & 2-channel, 100+ spk \\
\bottomrule
\end{tabular}
\caption{Representative public conversational corpora that full-duplex systems have used or could plausibly use for training. Total $\sim$5\,k\,h dominated by English telephone audio; Chinese-language public coverage is concentrated in MagicData-RAMC and HumDial Track\,2 (under 300\,h combined). Of these, Fisher is the empirically dominant training source --- used by dGSLM~\citep{nguyen2023generative} and Moshi~\citep{defossez2024moshi} --- with Easy Turn and HumDial Track\,2 the principal post-2024 additions.}
\label{tab:typec-public}
\end{table*}

Table~\ref{tab:typec-public} lists the principal public conversational corpora available for full-duplex training. Four observations: (i)~the LDC telephone era (Switchboard~\citep{godfrey1992switchboard}, CallHome, Fisher~\citep{cieri2004fisher}) remains the largest contiguous source of two-channel time-synchronous English audio; (ii)~multi-party meeting corpora (ICSI~\citep{janin2003icsi}, AMI~\citep{carletta2007ami}) are abundant in turn-taking phenomena but use multi-channel rather than two-speaker setups and are therefore better suited to diarization-flavored training; (iii)~the modern conversational era is best represented by CANDOR~\citep{reece2023candor} for English video-call audio and MagicData-RAMC~\citep{yang2022magicdata} for Mandarin mobile-phone audio; (iv)~the newest additions are explicitly turn-taking-aware: Easy Turn~\citep{li2026easy} is a large public turn-taking detection corpus with four-state labels (complete/incomplete/backchannel/wait) for end-of-turn classifier training, and HumDial Track\,2~\citep{wang2026humdial} provides a public corpus designed around full-duplex sub-scenarios (interruption, rejection, third-party speech, pause handling). The combined total of $\sim$5\,k\,h is substantially smaller than industrial Track-B totals (see \S\ref{app:data:tracks}), though the exact gap is not publicly verifiable from disclosed industrial documentation.

\subsection{Two Tracks: Public vs Industrial}
\label{app:data:tracks}

Full-duplex training data does not flow through a single shared pipeline. It splits structurally into two tracks with divergent evolution, accessibility, and coverage bias --- a split the rest of the field's data discourse should acknowledge explicitly.

\paragraph{Track A --- Public.} Driven by academic groups, public funding, and open-source releases; on the order of $\sim$1\,k\,h per release; mostly commercially usable; biased toward $T1$ / $T2$ / $T5$ cells (clean turns, latched handoffs, hesitation) with $T3$ at natural-distribution frequency and $T4$ essentially absent; updated on a five-to-ten-year cadence (Fisher 2003 $\to$ CANDOR 2020 $\to$ Easy Turn 2025).

\paragraph{Track B --- Industrial proprietary.} Continuously fed by deployed products (call centers, smart speakers, in-app voice). Scale and cell-coverage details are generally not disclosed; reports from individual deployments suggest larger volumes and domain-specific bias (e.g., call-center transcripts naturally over-represent some overlap patterns), but a quantitative comparison with Track A is not currently possible from public sources. Licensing is closed; reproducibility is zero.

The asymmetry of Track B over Track A on raw scale, combined with Track B's product-specific bias, has a load-bearing implication: an open-source full-duplex system trained only on Track-A data will, in the short term, neither match Track-B systems on overall behavior nor be auditable against them on cell-by-cell coverage. The data agenda in \S\ref{app:data:agenda} below is in large part an attempt to make Track A useful in spite of this gap.

\subsection{Synthesis Pipelines}
\label{app:data:synth}

Because real Type-C data is scarce and expensive, most modern systems augment with synthetic two-channel dialogues. We catalogue five recipes seen in the recent literature, in order of ontology-cell expressiveness:

\begin{enumerate}
  \setlength{\itemsep}{1pt}
  \item \textbf{Dual-role TTS}: an LLM generates a turn-by-turn script; a multi-speaker TTS synthesizes the two channels. Covers $T1$ only; clean turn-taking with no overlap.
  \item \textbf{Overlap injection}: dual-role TTS plus randomly injected onset overlaps at turn boundaries. Adds $T2$ (latched) and partial $T3 \cdot I4$ (barge-in). MinMo's heuristic $T \sim \mathcal{N}(0.6, 0.4^2)$ inter-turn distribution is an instance of this.
  \item \textbf{Backchannel insertion}: detects short back-channel utterances in source dialogue, randomly inserts them into the assistant's speaking span. Adds $T3 \cdot I2$ but requires source dialogue actually containing back-channels --- a circular dependency on Type-C data.
  \item \textbf{Silent-token augmentation}: OmniFlatten's~\citep{zhang2025omniflatten} \texttt{silent\_speech\_token} explicitly marks ``the system should be silent here'' in the flattened sequence, making $T5$ expressible at the token level without modifying the audio data.
  \item \textbf{Time-anchor self-supervision}: SyncLLM's~\citep{veluri2024beyond} \texttt{[S0]}/\texttt{[S1]} markers at chunk boundaries; teaches the model an explicit sense of within-chunk timing. A streaming-architecture support tool rather than a cell-expanding data augmentation.
\end{enumerate}

Two cells remain unreachable by any of these recipes alone: $T4$ (sustained concurrent speech), because synthesizing two coherent simultaneous monologues is itself unsolved; and $T3 \cdot I7$ (third-party / extraneous voice), because injecting a third speaker requires either real environmental recordings or a non-trivial pipeline.

\subsection{Ontology Cell $\times$ Training Corpus Coverage}
\label{app:data:matrix}

\begin{table*}[t]
\centering
\small
\setlength{\tabcolsep}{6pt}
\renewcommand{\arraystretch}{1.2}
\begin{tabular}{l ccccc}
\toprule
\textbf{Cell} & \textbf{Fisher} & \textbf{CANDOR} & \textbf{MagicData-RAMC} & \textbf{Easy Turn} & \textbf{HumDial T2} \\
\midrule
$T1 \cdot I1 \cdot R6$ Standard       & \checkmark & \checkmark & \checkmark & \checkmark & \checkmark \\
$T2 \cdot I1 \cdot R6$ Latched        & \checkmark & \checkmark & \checkmark & $\vartriangle$ & $\vartriangle$ \\
$T3 \cdot I4 \cdot R2$ Barge-in       & \checkmark & \checkmark & \checkmark & $\cdot$ & \checkmark \\
$T3 \cdot I2 \cdot R1$ Backchannel    & \checkmark & \checkmark & \checkmark & \checkmark & \checkmark \\
$T3 \cdot I7 \cdot R5$ Third-party    & $\cdot$ & $\cdot$ & $\cdot$ & $\cdot$ & $\vartriangle$ \\
$T5 \cdot I6 \cdot R3$ Hesitation     & \checkmark & \checkmark & \checkmark & \checkmark & \checkmark \\
$T4 \cdot I1 \cdot R1$ Concurrent     & $\cdot$ & $\cdot$ & $\cdot$ & $\cdot$ & $\cdot$ \\
\bottomrule
\end{tabular}
\caption{Ontology-cell coverage of representative public Type-C training corpora. \textbf{\checkmark}~well represented (Fisher, CANDOR, and MagicData-RAMC contribute $T3$ overlap cells at the natural conversational rate; Easy Turn contributes $T1$/$T3 \cdot I2$/$T5$ via designed turn classes; HumDial Track\,2~\citep{wang2026humdial} contributes $T3 \cdot I4$ and $T3 \cdot I2$/$I7$ via designed interruption/rejection sub-scenarios); \textbf{$\vartriangle$}~partial (CANDOR is a dyadic two-person video corpus with quality control excluding extra speakers; Easy Turn and HumDial cover $T2$ only as a degenerate end-of-turn case); \textbf{$\cdot$}~not represented for structural reasons (telephone and single-party-recording corpora structurally exclude third-party voice; no public corpus contains sustained concurrent monologue). HumDial Track\,2 includes designated $T3 \cdot I7$ third-party speech sub-tasks, though its third-party training subset is small (e.g., its FDBench train set has only $\sim$120 third-party items and 0 directed-to-others items per Table~1 of the challenge release). The FDB-v1.5 benchmark~\citep{lin2026full} is an evaluation resource and is treated separately in \S\ref{app:evaluation}.}
\label{tab:cell-dataset-coverage}
\end{table*}

Table~\ref{tab:cell-dataset-coverage} maps the seven ontology cells most discussed in this paper against the representative public Type-C \emph{training} corpora; we deliberately exclude the FDB-v1.5 benchmark from this matrix because it is an evaluation resource rather than a training source. Three patterns dominate:

\begin{itemize}
  \setlength{\itemsep}{1pt}
  \item \textbf{$T4$ (sustained concurrent) is the single most under-served cell across all public training data.} Combined with the equally thin evaluation coverage of $T4$ in current benchmarks (\S\ref{app:evaluation}), this explains why every system in \S\ref{app:systems} fails this cell.
  \item \textbf{$T3 \cdot I7$ (third-party voice) is newly addressed by HumDial Track\,2.} HumDial Track\,2~\citep{wang2026humdial} explicitly includes a Rejection scenario covering ``Third-party Speech'' and ``Speech Directed at Others'' as sub-tasks, although the training subsets for these categories remain small per its Table~1 (e.g., $\sim$120 third-party items and 0 directed-to-others items in the FDBench train split).
  \item \textbf{The CANDOR/MagicData/HumDial trio addresses the modern-acoustic / multilingual gap that Fisher and Switchboard leave.} A 2026 open-source FD training mix that combines Fisher (telephone, scale) with CANDOR (video-call acoustics), MagicData-RAMC (Mandarin), Easy Turn (turn-class labels), and HumDial Track\,2 (rejection sub-scenarios) covers six of the seven ontology cells; the remaining gap is $T4$.
\end{itemize}

\subsection{Data Research Directions}
\label{app:data:agenda}

The audit above motivates six concrete data-side research directions. We list them in approximate decreasing tractability for the open-source community:

\begin{enumerate}
  \setlength{\itemsep}{1pt}
  \item \textbf{Third-party / background voice synthesis (short-term)}: combine an existing single-stream voice library, multi-speaker TTS, and room-impulse-response convolution to produce $T3 \cdot I7$ training data. A pure engineering task; estimated person-month-scale to assemble a multi-thousand-hour synthetic corpus.
  \item \textbf{Ontology-driven synthesis pipeline (short-term)}: extend the five-recipe survey above into a single open tool with per-cell injection probabilities (e.g., $p_{T_2}, p_{T_3 \cdot I_4}, p_{T_3 \cdot I_2}, p_{T_3 \cdot I_7}, p_{T_4}$), so that any LLM-generated script can be expanded into two-channel dialogue with a target cell distribution rather than the natural distribution.
  \item \textbf{Cross-dataset alignment and combination (short-term)}: align Easy Turn ($T1$/$T5$), FDB-v1.5 ($T3$), and Fisher ($T1$/$T2$ + natural $T3$) under the ontology framework into a single ``ontology-typed'' open corpus.
  \item \textbf{$T_4$ targeted real-data collection (medium-term)}: design real-data collection protocols for sustained concurrent speech --- conference calls, group discussion, debate, family-room recordings. Pure synthesis cannot reach this cell at current state of the art; real collection is required.
  \item \textbf{Agent-vs-agent self-play data generation (medium-to-long-term)}: let two full-duplex models converse with each other under ontology-typed sampling and filtering. Risks amplifying biases of current models, so should be combined with the synthesis pipeline of direction~2 as an ontology constraint.
  \item \textbf{Training objectives that compensate for data scarcity (short-to-medium-term)}: recent work argues that better training objectives can substitute for missing data --- reinforcement learning with conversational rewards (e.g., dual-axis generative reward models that jointly score semantic quality and turn-taking robustness~\citep{chen2026dual}), or auxiliary turn-aware objectives that imprint turn structure on ordinary dialogue data. This is the only direction that does not require new data collection and is therefore the highest-leverage short-term direction for open-source teams.
\end{enumerate}

\subsection{Scope of the Thesis}
\label{app:data:scope}

The data-bottleneck thesis of \S\ref{sec:data} is a strong but bounded claim, and four boundaries are worth stating explicitly so that it is not over-read as ``architecture does not matter''.

\paragraph{The thesis applies only at L0--L2.} The hierarchy of \S\ref{sec:hierarchy-audit} has mature representatives at L0, L1, and L2. L3 representation-level full-duplex has no representative yet, and reaching it will require architectural innovations (continuous-latent autoregressive streaming, JEPA-style dual-stream prediction, world-model-conditioned dialogue) that are not data problems but modelling-paradigm problems.

\paragraph{Architecture and data are coupled, not independent.} The thesis does not claim that architecture choice has no effect on data efficiency. Multi-stream designs may convert $T3 \cdot I4$ data into capability more efficiently than flattened designs; channel-fusion designs may be more sample-efficient in low-data regimes. The thesis claims that the \emph{ceiling} of behavior is set by data; architecture sets the gain on the path from data to ceiling.

\paragraph{Engineering cost may favor the architecture path.} In an industrial setting, the cost of acquiring sufficient Type-C data may be comparable to or exceed the cost of architecture iteration. The thesis is a statement about research priorities, not about every team's optimal engineering choice.

\paragraph{Evaluation paradigms may shift the picture.} Current benchmarks (FDB family, MTR-Duplex) measure largely static turn-taking behaviors. If the field moves to multi-turn, agent-versus-examiner, adversarial evaluation, then systems with stronger reasoning traces (e.g., a Thinker--Talker decomposition) may demonstrate capability gaps that no amount of additional Type-C data closes. The thesis is best read as conditional on the current evaluation regime.

Within these bounds, the bottleneck thesis explains, with no further machinery, the cross-system performance variance reported in \S\ref{app:systems} and the open-source-versus-industrial gap that \S\ref{app:data:tracks} documents.

%% file: sections/07_evaluation.tex
\vspace{-2pt}

\section{Evaluation and the Realization Gap}
\label{sec:evaluation}
\label{app:evaluation}
\vspace{0pt}

Conventional speech metrics fail in the full-duplex regime by construction: WER measures the ASR module, MOS the TTS module, BLEU the response generator, and end-of-turn latency the scheduler --- each assumes that dialogue consists of discrete user-then-system events, and none answers the question that defines a full-duplex system, ``what should the system do per-frame while the user is still speaking?''. WER does not test graceful stopping when interrupted; MOS does not test floor-claim avoidance; end-of-turn latency presupposes a turn boundary that does not exist on the $T4$ concurrent cell. The benchmarks reviewed below are the response to this gap.

\paragraph{Two FDB series, not one.} We disambiguate two active benchmark lines with similar names. \textbf{Full-Duplex-Bench (FDB)}~\citep{lin2025fullv1,lin2026full,lin2025fullv2,lin2026fullv3} is a versioned suite (v1--v3) evolving from passive overlap tests to active multi-turn probes, notably reporting that audited systems often struggle with overlap, correction, and multi-turn entity tracking (mirroring our \S\ref{sec:data} data-bottleneck findings). Conversely, \textbf{FD-Bench}~\citep{peng2025fd} is an independent, configurable scenario generator. Other recent resources --- Talking Turns~\citep{arora2025talking}, MTR-Duplex~\citep{zhang2025mtr}, Easy Turn~\citep{li2026easy}, and the ICASSP 2026 HumDial challenge --- are detailed below alongside the per-benchmark cell-coverage matrix.

\paragraph{The realization gap.} The hierarchy of \S\ref{sec:hierarchy-audit} describes a system's \emph{architectural capacity} --- what cells the design admits in principle. The benchmarks above measure \emph{demonstrated behavior} --- what cells the trained system actually serves. The discrepancy is what we call the \emph{realization gap}, and it is the operative reading of the data-architecture-evaluation chain. The gap is empirically non-trivial: Mini-Omni2 and OmniFlatten both belong to the L2 flattened family with identical architectural reach, but the former demonstrates neither $T3 \cdot I2$ nor $T4$ on FDB while the latter claims both. The cause is downstream of architecture and upstream of evaluation --- it is the training-data cell coverage of \S\ref{sec:data}. We therefore recommend that every full-duplex system report a three-component profile: (i)~its L-layer, (ii)~the cell-coverage profile of its training data, and (iii)~its demonstrated per-cell behavior on the FDB v1.5 / v2 protocol. A single aggregate score collapses these three into a number that obscures the diagnostic question ``is the system failing because of architecture, data, or evaluation construct?''.

\paragraph{Structural gaps in the current regime.} Three regime-level gaps remain. First, almost all benchmark audio is static and pre-recorded: even FDB v2's examiner follows a scripted task family rather than a genuinely adaptive adversary. Second, public test audio is dominated by English under telephone or Zoom acoustics (a qualitative trend rather than a precisely measured share); industrial mobile-microphone and Mandarin conditions are not standardized. Third, among the public benchmarks we reviewed, none includes an explicit causal-influence probe distinguishing substantive from apparent full-duplex --- chunk-alternating designs can score well on stop-latency without truly integrating user audio mid-turn. We treat the third gap as the most actionable, and \S\ref{app:evaluation} sketches what such a probe would look like.

%% file: appendix/E_evaluation.tex

\subsection{Survey of Full-Duplex Benchmarks}
\label{app:eval:benchmarks}

\paragraph{Naming disambiguation.} The literature contains two independent, simultaneously active benchmark series with confusingly similar names. We name them explicitly here because cross-referencing between them is non-trivial and uncited mentions of ``FD-Bench'' in recent papers are sometimes ambiguous.

\begin{itemize}
  \setlength{\itemsep}{1pt}
  \item \textbf{Series~A --- Full-Duplex-Bench (FDB).} A versioned line by Lin et al.: v1~\citep{lin2025fullv1} $\to$ v1.5~\citep{lin2026full} $\to$ v2~\citep{lin2025fullv2} $\to$ v3~\citep{lin2026fullv3}, with v1.5 establishing the overlap-handling protocol most other recent work compares against. Focus: behavior-level evaluation, per-cell metrics.
  \item \textbf{Series~B --- FD-Bench.} An independent benchmarking pipeline by Peng et al.~\citep{peng2025fd} (data generator, test-case construction, scoring scripts) rather than a single fixed test set. Focus: system-level pipeline evaluation including throughput, latency, and robustness.
\end{itemize}

The two series are not interchangeable. They measure different things along different axes and their numbers are not directly comparable. Future work that needs to cite ``FD-Bench'' generically should explicitly name the team or the version.

\paragraph{Benchmark-by-benchmark notes.} We summarize the resources that we treat as load-bearing in this paper:

\begin{itemize}
  \setlength{\itemsep}{2pt}
  \item \textbf{Talking Turns}~\citep{arora2025talking}. Trains a supervised judge model on Switchboard human-human dialogue and uses it to score whether a full-duplex system's per-frame behavior (speak / pause / backchannel) matches the human distribution. Strength: anchored in real human turn-taking statistics. Limitation: covers $T1$, $T2$, and the $T3 \cdot I4$ slice of the ontology; silent on $T3 \cdot I2$ (backchannel-as-listener) and $T3 \cdot I7$ (third-party voice).
  \item \textbf{FDB v1}~\citep{lin2025fullv1} (Series~A). Four axes --- Pause, Backchannel, Turn-taking, Interruption --- with automated metrics for each. Coverage: $T5 \cdot I6 \cdot R3$, $T3 \cdot I2 \cdot R1$, $T1 \cdot I1 \cdot R6$, $T3 \cdot I4 \cdot R2$. Single-turn, English-dominant, static test set.
  \item \textbf{FDB v1.5}~\citep{lin2026full}. The key v1.5 contribution is the decomposition of overlap into four distinct scenarios --- user interruption, user backchannel, talking-to-others, background speech --- which align nearly one-to-one with the $T3$ row of the T\,$\times$\,I ontology. The \emph{stop-latency} metric (how quickly the system yields after a user onset) is, to our knowledge, the closest existing public metric to a substantive-FD test. A reported finding of v1.5 is that current systems split into two dominant strategies --- response-first (fast but easily confused by backchannels) and floor-holding (stable but slow on legitimate barge-in) --- which is itself a useful framing for the ontology audit of \S\ref{sec:ha:audit}.
  \item \textbf{FDB v2}~\citep{lin2025fullv2} (Series~A). Replaces the static test set with a multi-turn automated examiner that drives the system through four task families (daily / correction / entity-tracking / safety) at two paces (fast / slow). This is a paradigmatic shift from passive test set to active conversational partner. Reported finding: audited systems frequently struggle on the combination of concurrent speech ($T4$), mid-turn correction ($T3 \cdot I3$), and multi-turn entity tracking, though the paper reports these as qualitative often/sometimes patterns rather than uniform failures. The failure pattern coincides with the data-missing cells of \S\ref{app:data:matrix}, providing the most direct independent evidence we have for the data-bottleneck thesis.
  \item \textbf{FDB v3}~\citep{lin2026fullv3} (Series~A). Extends FDB to tool-using voice agents: it scores how a system handles tool calls under disfluent user behavior (hesitation, mid-call interruption, recovery). Treats the FD system as a component of a larger agent stack rather than as a dialogue partner in isolation; appropriate for voice-agent evaluation, less so for core FD architecture comparison.
  \item \textbf{MTR-Duplex}~\citep{zhang2025mtr}. Multi-round full-duplex evaluation focused on conversational content quality --- dialogue features, dialogue quality, instruction following, safety. Complementary to FDB v2: FDB v2 tests \emph{when} the system speaks, MTR-Duplex tests \emph{what} it says.
  \item \textbf{FD-Bench}~\citep{peng2025fd} (Series~B). Pipeline-style: provides a configurable generator for test scenarios rather than a fixed test set. Distinct in form and intent from Series~A; appropriate when evaluating engineering robustness rather than behavioral coverage.
  \item \textbf{Easy Turn test set}~\citep{li2026easy}. The test slice of the Easy Turn corpus measures four-class turn classification: complete / incomplete / backchannel / wait. It is a large public open-source resource with four-state labels and a published baseline that enables comparable turn-class accuracy reporting across systems.
  \item \textbf{HumDial Challenge 2026}~\citep{wang2026humdial}. Announced as releasing (i)~a dual-channel real human-dialogue corpus, (ii)~a public leaderboard, and (iii)~a standardized evaluation protocol. The signature open-evaluation event of 2026 for full-duplex.
\end{itemize}

\paragraph{Survey companions.} \citet{ji2024wavchat} survey end-to-end spoken dialogue models more broadly; their evaluation chapter overlaps in scope with this section but predates the FDB v1.5 / v2 wave. \citet{chen2025turn} review turn-taking specifically and provide useful pre-FDB context.

\subsection{Ontology-Cell Coverage of Benchmarks}
\label{app:eval:coverage}

\begin{table*}[t]
\centering
\small
\setlength{\tabcolsep}{6pt}
\renewcommand{\arraystretch}{1.2}
\begin{tabular}{l ccccc}
\toprule
\textbf{Cell} & \textbf{Talking Turns} & \textbf{FDB v1} & \textbf{FDB v1.5} & \textbf{FDB v2} & \textbf{Easy Turn} \\
\midrule
$T1 \cdot I1 \cdot R6$ Standard       & \checkmark & \checkmark & \checkmark & \checkmark & \checkmark \\
$T2 \cdot I1 \cdot R6$ Latched        & \checkmark & $\vartriangle$ & $\vartriangle$ & \checkmark & $\vartriangle$ \\
$T3 \cdot I4 \cdot R2$ Barge-in       & \checkmark & \checkmark & \checkmark & \checkmark & $\cdot$ \\
$T3 \cdot I2 \cdot R1$ Backchannel    & $\cdot$    & \checkmark & \checkmark & $\vartriangle$ & \checkmark \\
$T3 \cdot I7 \cdot R5$ Third-party    & $\cdot$    & $\cdot$    & \checkmark & $\vartriangle$ & $\cdot$ \\
$T5 \cdot I6 \cdot R3$ Hesitation     & $\vartriangle$ & \checkmark & $\vartriangle$ & \checkmark & \checkmark \\
$T4 \cdot I1 \cdot R1$ Concurrent     & $\cdot$    & $\cdot$    & $\cdot$    & $\vartriangle$ & $\cdot$ \\
\bottomrule
\end{tabular}
\caption{Ontology-cell coverage of the principal evaluation resources. \textbf{\checkmark}~the cell is directly tested by an explicit metric or sub-task in the benchmark; \textbf{$\vartriangle$}~the cell is partially tested, either as a subset of a broader category or through a metric that does not fully isolate it; \textbf{$\cdot$}~the cell is not tested. FDB v1.5's row reflects its overlap-into-four-scenarios decomposition, including \emph{talking-to-others} / \emph{background speech} which together cover $T3 \cdot I7$. FDB v2 partially reaches $T4$ via its examiner's concurrent-speech probes. No public benchmark currently tests $T4$ as a primary metric; this is the single largest hole in the evaluation landscape.}
\label{tab:eval-cell-coverage}
\end{table*}

Table~\ref{tab:eval-cell-coverage} maps the principal benchmarks against the same seven ontology cells used in \S\ref{app:data:matrix}. Three patterns are visible:

\begin{itemize}
  \setlength{\itemsep}{1pt}
  \item \textbf{$T1$ Standard and $T3 \cdot I4$ Barge-in are over-served} relative to other cells: every benchmark in the table covers both, often with multiple metrics each. Comparing systems on these cells alone is unlikely to yield new information.
  \item \textbf{$T4$ Concurrent remains under-served on the evaluation side too} --- only FDB v2's examiner partially probes it. The training-data scarcity for $T4$ documented in \S\ref{app:data:matrix} is therefore mirrored on the evaluation side: even teams that solve the data problem would currently lack a clean way to demonstrate the solution.
  \item \textbf{FDB v1.5 is the single most useful benchmark for the $T3$ row} by virtue of its overlap-scenario decomposition. Its \emph{stop-latency} metric is the closest existing operationalization of the substantive-FD criterion of \S\ref{sec:state-machine}, and is the metric the rest of this section recommends as the empirical anchor.
\end{itemize}

\subsection{The Realization Gap}
\label{app:eval:realization}

The hierarchy of \S\ref{sec:hierarchy-audit} describes a system's \emph{architectural capacity} --- what cells the design admits in principle. The benchmarks of \S\ref{app:eval:benchmarks} measure the system's \emph{demonstrated behavior} --- what cells the trained system actually serves on the test set. The discrepancy between the two is what we call the \emph{realization gap}, and it is the conceptual contribution this section recommends taking forward.

The realization gap is empirically non-trivial. Two L2 systems with the same architectural capacity can have very different demonstrated behavior on the same cell --- Mini-Omni2 and OmniFlatten both belong to the L2 flattened family, but the former demonstrates neither backchannel nor concurrent capability on FDB while the latter claims both. The cause is downstream of architecture and upstream of evaluation: it is the cell-coverage distribution of training data, the subject of \S\ref{app:data}. The realization gap thus operationalizes the link between \S\ref{app:systems} and \S\ref{app:data}: \emph{architecture sets the ceiling, training data realizes some fraction of it, and evaluation reports that fraction}.

A useful evaluation report for a full-duplex system therefore has three components, not one: (i)~the architectural L-layer of the system (from \S\ref{app:systems}); (ii)~the cell-coverage profile of its training data (from \S\ref{app:data}); and (iii)~the demonstrated per-cell behavior on the benchmarks of \S\ref{app:eval:benchmarks}. A single aggregate score collapses these three into a number that obscures the diagnostic question --- ``is this system failing because of architecture, data, or evaluation construct?'' --- that an evaluation should answer.

This is our conceptual recommendation. We do not propose new metrics or a new benchmark; we propose that existing benchmarks be reported under this three-component frame.

\subsection{Structural Gaps in the Current Regime}
\label{app:eval:gaps}

We close with three structural gaps that the current evaluation regime exhibits. None of them is fully closable in the short term, but each suggests a concrete direction for the next round of benchmarks.

\paragraph{Static test sets versus genuinely adaptive interaction.} All public FDB versions before v2, all of Easy Turn, and all of Series~B's pipeline scenarios are pre-recorded: the system's response is logged but does not change what the test set asks next. Real conversation is recursive in exactly the opposite direction --- the human partner adapts to what the system just did. FDB v2's automated examiner is a partial fix but the examiner itself follows a scripted task family rather than mounting a genuinely adaptive adversary.

\paragraph{Language and acoustic distribution.} Public benchmark audio is dominated by English under telephone or Zoom-style acoustic conditions; we did not find duration-by-language metadata sufficient to give a precise share, but the qualitative bias is consistent across the benchmarks we reviewed. Industrial systems serve mobile-microphone audio in noisy environments in Chinese and other languages, but there is no standardized way to report results in those conditions. The result is that the open-source / industrial gap discussed in \S\ref{app:data:tracks} is reinforced rather than measured by the current benchmarks.

\paragraph{Absence of explicit substantive-versus-apparent probes.} No public benchmark currently includes a causal-influence test of the kind described in \S\ref{sec:state-machine} --- a controlled counterfactual experiment that asks whether the system's mid-turn behavior is genuinely sensitive to the user audio, or whether it is following a fixed schedule. Chunk-alternating designs (e.g., the SyncLLM family) can score well on stop-latency without passing such a test, and the field would benefit from a benchmark that explicitly distinguishes the two.

Taken together, the three gaps suggest that the next round of evaluation work should aim for an adaptive multi-turn protocol with explicit causal-influence probes and multilingual / multi-acoustic test conditions. None of these directions requires inventing new metrics; they require recombining the protocols already represented in the benchmarks of \S\ref{app:eval:benchmarks} with the data agenda of \S\ref{app:data:agenda}.

%% file: sections/08_frontiers_conclusion.tex
\section{Frontiers and Conclusion}
\label{sec:frontiers-conclusion}

Sections \S\ref{sec:data} and \S\ref{sec:evaluation} identify $T4$ sustained concurrent speech as the primary residual gap in the L0--L2 frame, unserved by current models due to its absence from public corpora and benchmark metrics. While this gap is theoretically closable by data alone, a deeper architectural frontier remains: the L3 row of Table~\ref{tab:audit} is empty. We treat this as the field's main forward-looking hypothesis.


\subsection{The L3 Hypothesis}
\label{sec:fc:l3}

L3 would dissolve the token-level boundary between perception and generation and make the \textsc{Dual} state native to a shared latent representation rather than negotiated through discrete codebooks. Three pieces of upstream technical evidence make this concrete enough to be a hypothesis rather than a placeholder, and we list three candidate architectures the survey treats as live options.

\paragraph{(i) Continuous-latent autoregressive streaming.} MAR~\citep{li2024autoregressive} shows an autoregressive Transformer can predict continuous-valued tokens via a small diffusion/flow-matching head~\citep{lipman2022flow}, removing the K-tokens-per-frame bottleneck of every L2 sub-design (\S\ref{app:found:rvq}). Transposed to dialogue, user and assistant audio would share a continuous-latent stream and \textsc{Dual} would be ``two streams active'' in the same latent --- not a separate mechanism. Kimi-Audio's streaming flow-matching detokenizer~\citep{ding2025kimi} and the codec-free SALMONN-omni~\citep{yu2026salmonn} are partial synthesis-side precursors.

\paragraph{(ii) JEPA-style dual-stream prediction.} The JEPA family~\citep{assran2023ijepa,bardes2024vjepa} learns representations by predicting masked latent regions from visible context rather than reconstructing raw signals. A dual-stream audio JEPA --- assistant latent predicted from joint user-plus-assistant context --- would natively integrate $T3 \cdot I*$ overlap rather than decide stop/yield on top of a single-speaker stream, and would inherit the self-supervised property that the model learns full-duplex dynamics from unlabeled two-channel audio without explicit turn-taking annotations.

\paragraph{(iii) World-model-conditioned dialogue.} Generative-world-model work~\citep{bruce2024genie} learns a latent encoding the controllable state of the world rather than the surface signal. Applied to dialogue, the latent would encode the dialogue state (who holds the floor, what was just said, what is expected next) and the audio surface would be conditionally generated; \textsc{Idle}/\textsc{Listen}/\textsc{Speak}/\textsc{Wait}/\textsc{Dual} of \S\ref{sec:state-machine} would correspond to latent regions, not external state variables.

\paragraph{Obstacles.} Three obstacles keep all three unrealized: (a)~the discrete-tokenizer ecosystem is now a deep substrate that continuous-latent designs must displace; (b)~streaming flow-matching and continuous-AR are not yet real-time at the 200--500\,ms latency current L2 systems achieve; (c)~no agreed formulation of what an L3 dialogue latent should encode exists, so the three hypotheses remain candidate framings rather than unified theory.
\subsection*{Conclusion}

This survey contributes three complementary frameworks (L0--L3 hierarchy, T$\times$I$\times$R ontology, five-state machine) grounded by three empirical audits (architecture, data, evaluation). Together, they replace the binary ``is this full-duplex?'' question with structured profiles that situate systems, align benchmarks, and expose specific gaps like the realization gap. We hope this shared vocabulary brings clarity to the recent proliferation of systems (\S\ref{sec:history}), providing a concrete starting point for next-generation corpora, evaluations, and L3 architectures.



\section*{Limitations}

Two limitations bound our claims. First, our per-system and cell-coverage audits rely on self-reported behaviors rather than independent end-to-end reproductions, compounded by the fact that several industrial L1 systems are closed-source. Second, our frameworks are calibrated against existing L0--L2 architectures. What a genuine L3 system entails remains open, and our candidate architectures (\S\ref{sec:fc:l3}) are provisional extrapolations. If L3 ultimately introduces novel interaction paradigms, our ontology and state machine will need extension rather than mere application.